\documentclass[a4paper,pra,twocolumn,10pt,floatfix,accepted=2023-05-09]{quantumarticle}
\pdfoutput=1 
\usepackage{amssymb,amsmath,amstext}
\usepackage{graphicx}
\usepackage{epstopdf}
\usepackage{color}
\usepackage{bm}
\usepackage[T1]{fontenc}
\usepackage{bbold}
\usepackage{bbm}
\usepackage{latexsym}
\usepackage{braket}
\usepackage{diagbox}
\usepackage[numbers]{natbib}
\usepackage{listings}
\usepackage{hyperref}
\usepackage[resetlabels]{multibib}

\newcommand{\tr}{{\rm Tr}}
\renewcommand{\vec}[1]{\boldsymbol{#1}}  
\usepackage{amssymb}% http://ctan.org/pkg/amssymb
\usepackage{pifont}% http://ctan.org/pkg/pifont
\newcommand{\cmark}{\ding{51}}%
\newcommand{\xmark}{\ding{55}}%

\begin{document}

\title{Unifying and benchmarking state-of-the-art quantum error mitigation techniques}

\author{Daniel Bultrini}
\thanks{The first two authors contributed equally to this work.}
\affiliation{Theoretical Division, Los Alamos National Laboratory, Los Alamos, NM 87545, USA}
\affiliation{Theoretische Chemie, Physikalisch-Chemisches Institut, Universität Heidelberg, INF 229, D-69120 Heidelberg, Germany}
\author{Max Hunter Gordon}
\thanks{The first two authors contributed equally to this work.}
\affiliation{Instituto de Física Teórica, UAM/CSIC, Universidad Autónoma de Madrid, Madrid, Spain}

\author{Piotr Czarnik}
\affiliation{Theoretical Division, Los Alamos National Laboratory, Los Alamos, NM 87545, USA}
\affiliation{Institute of Theoretical Physics, Jagiellonian University, Krakow, Poland.}

\author{Andrew Arrasmith}
\affiliation{Theoretical Division, Los Alamos National Laboratory, Los Alamos, NM 87545, USA}
\affiliation{Quantum Science Center, Oak Ridge, TN 37931, USA}

\author{M. Cerezo}
\affiliation{Information Sciences, Los Alamos National Laboratory, Los Alamos, NM 87545, USA}
\affiliation{Quantum Science Center, Oak Ridge, TN 37931, USA}

\author{Patrick J. Coles}
\affiliation{Theoretical Division, Los Alamos National Laboratory, Los Alamos, NM 87545, USA}
\affiliation{Quantum Science Center, Oak Ridge, TN 37931, USA}

\author{Lukasz Cincio}
\affiliation{Theoretical Division, Los Alamos National Laboratory, Los Alamos, NM 87545, USA}
\affiliation{Quantum Science Center, Oak Ridge, TN 37931, USA}

\begin{abstract}
 
Error mitigation is an essential component of achieving a practical quantum advantage in the near term, and a number of different approaches have been proposed. In this work, we recognize that many state-of-the-art error mitigation methods share a common feature: they are data-driven, employing classical data obtained from runs of different quantum circuits. For example, Zero-noise extrapolation (ZNE) uses variable noise data and Clifford-data regression (CDR) uses data from near-Clifford circuits. We show that Virtual Distillation (VD) can be viewed in a similar manner by considering classical data produced from different numbers of state preparations. Observing this fact allows us to unify these three methods under a general data-driven error mitigation framework that we call UNIfied Technique for Error mitigation with Data (UNITED). In certain situations, we find that our UNITED method can outperform the individual methods (i.e., the whole is better than the individual parts). Specifically, we employ a realistic noise model obtained from a trapped ion quantum computer to benchmark UNITED, as well as other state-of-the-art methods, in mitigating observables produced from random quantum circuits and the Quantum Alternating Operator Ansatz (QAOA) applied to Max-Cut problems with various numbers of qubits, circuit depths and total numbers of shots. We find that the performance of different techniques depends strongly on  shot budgets, with more powerful methods requiring more shots for optimal performance.  For our largest considered shot budget ($10^{10}$), we find that UNITED gives the most accurate mitigation. Hence, our work represents a benchmarking of current error mitigation methods and provides a guide for the regimes when certain methods are most useful.

\end{abstract}
\maketitle

\section{Introduction} 

As new generations of quantum computers become larger and less noisy, the practical application of quantum algorithms is just around the corner. It is likely that any quantum algorithm that can achieve an advantage over its classical counterpart will have to contend with substantial hardware errors. While quantum error correction may eventually allow one to eliminate hardware noise, doing so will require many more qubits and significantly lower error rates. Although full error correction will not be practical in the near term, it is possible to partially mitigate errors with far lower resource requirements. A number of such error mitigation techniques have been proposed~\cite{endo2021hybrid,cerezo2020variationalreview}, each with its own benefits and trade-offs. 

\begin{figure}[h]
    \centering
    \includegraphics[width = \columnwidth]{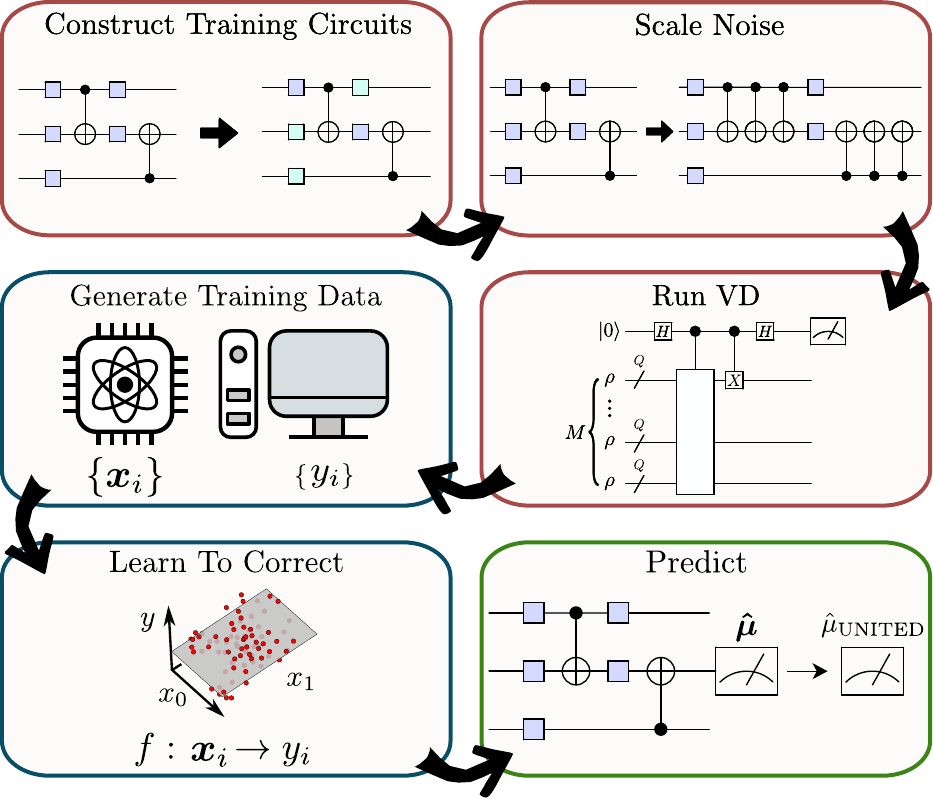}
    \caption{\textbf{Schematic summary of UNIfied Technique for Error mitigation with Data.} For a given circuit of interest near-Clifford training quantum circuits similar to the mitigated circuit are constructed.   For these circuits, expectation values of the observable of interest are obtained using a noisy quantum computer of interest for multiple noise strengths. Furthermore, Virtual Distillation (VD) mitigated expectation values are obtained for different numbers of state copies. The expectation values of the training circuits are also evaluated classically. These expectation values are then used to fit an ansatz that maps the noisy and VD-mitigated expectation values to the exact expectation value. The trained ansatz is then applied to analogous noise-scaled and VD-mitigated expectation values produced by the circuit of interest to mitigate the effects of the noise improving on the VD-mitigated observables.}
    \label{fig:summary}
\end{figure}

One of the most popular approaches to error mitigation is zero noise extrapolation (ZNE)~\cite{temme2017error}. ZNE involves collecting data at different noise levels and fitting an extrapolation to the noiseless case. A variety of methods have been introduced to amplify the noise as well as to perform the extrapolation~\cite{temme2017error,endo2018practical}. ZNE has been shown to work well for small problem sizes~\cite{kandala2018error} but has faced challenges for larger problems~\cite{czarnik2020error}. Due to the uncertainty of the extrapolation, ZNE does not have performance guarantees for regimes of high noise. This is particularly noticeable for circuits involving many gates. Here, the lowest available error points are likely to be too noisy for such fits to be helpful.

Alternatively, classically simulable circuits can be used to inform the experimenter about noise in a quantum device~\cite{czarnik2020error,strikis2020learning, montanaro2021error, vovrosh2021simple, urbanek2021mitigating}. The Clifford Data Regression (CDR) method~\cite{czarnik2020error} makes use of the Gottesman-Knill theorem~\cite{gottesman1998heisenberg} which guarantees that quantum circuits containing only Clifford gates can be classically simulated. CDR constructs a training set using data obtained from classically simulable near-Clifford circuits similar to the generic quantum circuit of interest. This set of data is then used to perform a regressive fit on an ansatz mapping noisy to exact expectation values. The fitted ansatz is used to estimate the value of the noiseless result. Furthermore, the combination of CDR and ZNE, which can outperform either method separately, has been considered and is called variable-noise CDR (vnCDR)~\cite{lowe2020unified}. vnCDR uses training data involving near-Clifford training circuits evaluated at multiple noise levels and can be thought of as using Clifford data to inform the extrapolation to the zero noise limit.

More recently, error mitigation via virtual distillation (VD) has been introduced~\cite{koczor2020exponential,huggins2020virtual}. VD mitigates noise by simultaneously preparing multiple copies of a noisy state, entangling them, and making measurements to virtually prepare a state which is purer than the original noisy states. Further improvements have been made and analyses have been performed~\cite{czarnik2021qubit,koczor2021dominant,huo2021dual, cai2021resourceefficient}. While VD could achieve exponential error suppression in some regime~\cite{koczor2020exponential,huggins2020virtual,czarnik2021qubit}, it has been noted that VD requires a number of shots that scales exponentially in the number of qubits used for sufficiently large noise rates~\cite{czarnik2021qubit}.

Given the success of merging ZNE and CDR into vnCDR, in this paper, we consider whether or not it is beneficial to unify these other error mitigation methods with VD. By considering the number of copies used in VD to be a noise-control parameter, we use Clifford data to guide a combination of VD-mitigated observables, much like vnCDR. While also adding ZNE-like noise amplification we arrive at a framework unifying CDR, ZNE, and VD, which we call UNIfied Technique for Error mitigation with Data (UNITED). A summary of the steps used in UNITED is shown in Figure~\ref{fig:summary}.

In order to get a realistic idea of how ZNE, vnCDR, VD, and UNITED compare we carry out a numerical study benchmarking the performance of each technique when mitigating a local observable produced by a random quantum circuit, and  one  term of a Max-Cut problem cost function for a circuit obtained when optimizing the Quantum Alternating Operator Ansatz (QAOA). We explore their performance on a simulator based on the current error rates of an ion-trap quantum computer. We compare performance across various system sizes and circuit depths, involving systems of up to $10$ qubits and depths of up to $128$ Molmer-Sørensøn entangling gate layers. Furthermore, we systematically compare the necessary shot cost for each method in its mitigation task, providing a fair comparison to guide experimenters in this field. 

Overall, for the largest considered shot budget ($10^{10}$) our new unified method UNITED gives an advantage over all other methods considered in mitigating observables produced by both random quantum circuits (RQC) and QAOA applied to the Max-Cut problem. We observe that ZNE outperforms the learning-based methods for the $10^5$ shot budget but fails to improve significantly when the total number of shots used is increased. This is in contrast to the learning-based methods, such as vnCDR and UNITED. These outperform ZNE when using $10^{6}$ and $10^{8}$ shots respectively.  VD performs similarly to ZNE in the RQC task but performs significantly better in the QAOA task being the best method for $10^5, 10^6$, and $10^8$ shot budgets. That is expected, as solutions of the Max-Cut problems are eigenstates of each term in the cost function which is beneficial to the performance of VD~\cite{koczor2021dominant}. Nevertheless, for the highest shot budget we consider ($10^{10}$), UNITED improves on the best-performing mitigation methods for QAOA and the RQC experiments. These results suggest that the unified method successfully leverages the strengths of each of its constituent parts. Hence, we believe that is it crucial to explore such unified methods.

\section{State-of-the-art error mitigation methods} 

\subsection{Zero Noise Extrapolation }
ZNE~\cite{temme2017error} uses noisy expectation values of an observable of interest $X$ obtained from the quantum circuit of interest $U$ run at increased noise to extrapolate to the noise-free limit. Originally,  Richardson extrapolation~\cite{richardson1927deferred} was proposed to estimate the noise-free observable~\cite{temme2017error}. The method was applied successfully to mitigate variational ground state quantum simulations with IBM quantum computers~\cite{kandala2018error, dumitrescu2018cloud}. 

Let us now explain the ZNE setting more formally. Here, the goal is to estimate the noiseless expectation value  $\mu = \bra{0}^{\otimes Q}U^\dag X U \ket{0}^{\otimes Q}$ from measurements on a noisy quantum device, where  $\ket{0}^{\otimes Q}$ denotes the input state to the quantum circuit. In a realistic scenario, one cannot experimentally access $\mu$ due to the presence of hardware noise. Instead, one measures in the quantum computer the noisy observable $\hat{\mu}$. In the ZNE framework one usually assumes that the noise in the device can be characterized by a single parameter $\lambda$ which is the noise strength. Moreover, while we assume that the noise cannot be decreased past a certain value $\lambda_0$, there always exist ways to further increase the noise strength (see below). Thus, in ZNE one wishes to estimate the noisy observable for increased noise levels, and then leverage extrapolation techniques to obtain the noiseless value. To perform the extrapolation, we measure the observable  for $n+1$ noise strengths  $\lambda_j = c_j \lambda_0$, where $1 = c_0 < c_1 < \dots < c_{n}$ \ coefficients are noise levels. This allows us to create a dataset composed of  expectation values estimated at different noise levels   $\mathcal{T}^{{\rm ZNE}}=\{\hat{\mu}_{j}^{\rm ZNE} \}_{j=0}^n$. The Richardson extrapolation estimates the  noise-free value $\mu$ via $\mathcal{T}^{{\rm ZNE}}$ as 
\begin{equation}
\mu^{{\rm ZNE}}_n = \sum_{j=0}^{n} \gamma_j \hat{\mu}^{\rm ZNE}_{j}, 
\label{Richardson}
\end{equation}
where 
\begin{equation}
\sum_{j=0}^{n} \gamma_j =1, \quad \sum_{j=0}^{n} \gamma_j c_j^k = 0 \, \ {\rm for} \, k=1,\ldots,n.
\label{eq:Richardson_coeff}
\end{equation}

 Under the standard assumption, Richardson extrapolation approximates the true noiseless value $\mu$ with an error scaling as $|\mu^{{\rm ZNE}}-\mu|\in\mathcal{O}(\lambda_{0}^{n+1})$~\cite{temme2017error}, see Appendix~\ref{app:Richardson} for details.  Nevertheless, the Richardson extrapolation in the presence of shot noise leads to the exponential growth of the uncertainty in $\mu^{{\rm ZNE}}$ with increasing $n$~\cite{giurgica2020digital}.
Richardson extrapolation is equivalent  to  fitting a polynomial of order $n$  on the various noisy expectation values, treating the noise level $c_j$ as an independent variable~\cite{he2020zero}. To address problems with the reliability of Richardson extrapolation for large $n$, lower order polynomial fits and exponential fits have been proposed~\cite{dumitrescu2018cloud,giurgica2020digital,cai2020multi}. Furthermore, approaches for  noise characterized  by multiple parameters have been postulated~\cite{otten2019recovering}.   

Regarding how one can increase the noise strength, the original proposal stretches pulses used to implement gates to increase the hardware noise~\cite{temme2017error}. Nevertheless, it is often more convenient to increase noise performing so-called identity insertions~\cite{dumitrescu2018cloud,he2020zero, giurgica2020digital}. These methods insert groups of gates 
equivalent to the identity into the circuit of interest.
 This does not change a state encoded by the noiseless  circuit  $U$ but increases the number of gates in the circuit, scaling the noise. 

ZNE provides a very simple framework to mitigate errors introduced by noise on some observable of interest. However, there are several technical challenges that limit its power. Firstly, ZNE assumes that low enough noise levels are experimentally accessible to enable successful extrapolation to the zero noise limit. Furthermore, the best choice of the ansatz used to  perform the extrapolation is not known a priori~\cite{giurgica2020digital}.  Also, a controlled  increase of the  noise strength is hardware dependent and may be challenging to implement in practice~\cite{kandala2018error}. 

\subsection{Clifford Data Regression }
Let us now introduce the CDR framework which provides background for the more  general vnCDR and UNITED methods. CDR~\cite{czarnik2020error} uses classically simulable training circuits consisting mainly of Clifford gates to learn the effects of the noise on an observable of interest. In particular, given the circuit of interest $U$, one constructs a set of circuits $\{V_i\}_{i=1}^{N_t}$, which are chosen to be similar to $U$. The simplest strategy to construct such a training set is to replace most of  the non-Clifford gates in the circuit of interest with  Clifford gates~\cite{czarnik2020error}. Here we note that to ensure classical simulability of the training circuits, they contain a small number $N$ of non-Clifford gates. This is due to the fact that  the cost of their classical simulation grows exponentially with $N$.

For each training circuit one computes via classical simulations~\cite{gottesman1998heisenberg}  the noiseless expectation value of the observable of interest $\mu_i = \bra{0}^{\otimes Q}V_i^\dag X V_i \ket{0}^{\otimes Q}$, as well as uses a quantum computer to estimate the noisy expectation value $\hat{\mu}^{{\rm CDR}}_i$ of $X$. Here, we will denote as $\mathcal{E}_{\rm noiseless}$ and $\mathcal{E}_{\rm noisy}$ the space of noiseless and noisy expectation values, respectively, so that $\mu_i\in \mathcal{E}_{\rm noiseless}$ and $\hat{\mu}_i^{{\rm CDR}}\in\mathcal{E}_{\rm noisy} $. Then, a dataset  $\mathcal{T}^{{\rm CDR}}=\{(\mu_i,\hat{\mu}_i^{{\rm CDR}})\}_{i=1}^{N_t}$ is used to train a  function $f^{\rm CDR}:\mathcal{E}_{\rm noisy}\rightarrow\mathcal{E}_{\rm noiseless}$, i.e.,  a function mapping the noisy to the exact expectation values. The simplest form for $f^{\rm CDR}$ which approximates a relation between the noisy and exact expectation values is
\begin{align}\label{eqn:cdr_ansatz}
    f^{\rm CDR}(x) = a_1 x+ a_2\,,
\end{align}
Here, the values of parameters $a_1, a_2$ are  found  by least-squares linear  regression on  $\mathcal{T}^{\rm CDR}$. Once the function is trained, CDR estimates the noiseless expectation value for $U$ as $\mu^{{\rm CDR}}=f^{\rm CDR}(\hat{\mu})$.
This method has been successfully applied to  quantum simulations of condensed matter systems with IBM quantum computers~\cite{czarnik2020error,sopena2021simulating}.     

One feature of near-Clifford circuits constructed in this manner is the clustering of their exact expectation values around $0$. This can lead to a large clustering of the expectation values in a near-Clifford training set which is detrimental to the shot efficiency of such an approach \cite{czarnik2022improving}. Therefore, to improve the shot efficiency, strategies have been proposed to ensure sufficient variance of the exact observable of interest among the training circuits \cite{czarnik2022improving}.
Alternatively, the usage of Markov Chain Monte Carlo was proposed to construct training sets, where the training circuits have a low value of a cost function in the case of variational quantum algorithms~\cite{czarnik2020error}.

The primary challenge in CDR is the construction of the near-Clifford training circuits as the optimal procedure to construct the training set is not known in general. Furthermore, the method  requires evaluation of the observable of interest for the training circuits which may result in a large shot cost of the mitigation for large training sets. 

\subsection{Variable noise Clifford Data Regression }
Variable noise Clifford Data Regression (vnCDR) conceptually unifies ZNE with CDR by using a training set involving near-Clifford circuits evaluated at several noise levels~\cite{lowe2020unified}. The dataset for vnCDR is constructed as follows. First, just like in CDR, one  constructs a dataset of circuits $\{V_i\}_{i=1}^{N_t}$ which are chosen to be similar to $U$. Subsequently, one uses a classical computer to obtain the noiseless expectation value of the observable of interest $\mu_i$. Then, for each $V_i$ and each noise level  $c_{j}$ one uses a quantum computer to estimate the noisy expectation values $\hat{\mu}^{{\rm vnCDR}}_{i,j}$, just like in ZNE. This allows one to obtain a  training dataset of the form $\mathcal{T}^{\rm vnCDR} = \{(\hat{\vec{\mu}}_i^{{\rm vnCDR}}, \mu_i)\}_{i=1}^{N_t}$, where $\hat{\vec{\mu}}_i^{{\rm vnCDR}} = (\hat{\mu}^{{\rm vnCDR}}_{i,0},\dots,\hat{\mu}^{{\rm vnCDR}}_{i,n})$. We note that for $n=0$ one recovers the CDR training set  $\mathcal{T}^{{\rm CDR}}$. The dataset $\mathcal{T}^{\rm vnCDR}$ is then used to fit a linear function $f^{\rm vnCDR}:(\mathcal{E}_{\rm noisy})^{ (n+1)}\rightarrow\mathcal{E}_{\rm noiseless}$, of the form
\begin{equation}
       f^{\rm vnCDR}(\vec{x}) = \sum_{j=0}^{n} a_j x_{j}\,.
\label{eq:vnCDR_ansatz}
\end{equation}
This function is trained using least-squares regression which leads to the best-fit parameters $a^{*}_{j}$. These fitted parameters $a^{*}_{j}$ are then used to obtain the vnCDR estimate of $\mu$, as  $\mu^{\rm vnCDR}=f^{\rm vnCDR}(\vec{\hat{\mu}}^{\rm ZNE})$. Here, we have defined $\hat{\vec{\mu}}^{\rm ZNE} = (\hat{\mu}_0^{\rm ZNE},\dots,\hat{\mu}_n^{\rm ZNE})$ as the vector  comprised of the $n+1$ noisy expectations for the circuit of  interest (that is, $\hat{\vec{\mu}}^{\rm ZNE}$ is a vector of the data in the ZNE dataset $ \mathcal{T}^{{\rm ZNE}}$). The vnCDR method has been successfully implemented for quantum dynamics simulations with IBM quantum computers~\cite{sopena2021simulating}.

As we can see, vnCDR shares several common features with CDR and ZNE. The functional form of the ansatz is a linear combination of noisy estimates which resembles Richardson extrapolation.  Therefore, vnCDR can be understood as performing an extrapolation guided by near-Clifford circuits with a similar structure to the circuit of interest. Nevertheless, the fitted ansatz does not require explicit knowledge of the noise levels, unlike the ZNE ansatzes, potentially easing ZNE requirements for precise control of the noise strength.  Both CDR and vnCDR use training on data produced by near-Clifford quantum circuits. Consequently, vnCDR can also be treated as adding additional relevant data features to the CDR training set. We note that the ansatz for vnCDR does not include a constant term unlike that of CDR. 

As in the case of CDR, the construction of an informative training set is the primary technical challenge. Furthermore, adding additional noise levels to the vnCDR training set increases the shot cost of the mitigation. 
Therefore, the best choice of noise levels is not known a priori.

\subsection{Virtual Distillation }

Virtual distillation (VD)~\cite{koczor2020exponential,huggins2020virtual} takes a slightly different approach  than ZNE and CDR. While the former suppresses error via post-processing, VD attempts to mitigate the effect of noise coherently. Specifically, let us denote by $\rho$ the noisy state obtained at the output of the circuit of interest $U$. In VD, one 
uses $M>1$ copies of a noisy state $\rho$  to obtain  expectation values of an  observable of interest $X$ for a purer, ``distilled'' state. Specifically,   VD mitigates the  expectation value of $X$ by computing  the quantity 
\begin{equation}
\hat{\mu}^{\rm VD} = \frac{{\rm Tr } [\rho^M X]}{{\rm Tr} [\rho^M]}.  
\end{equation}

To better understand VD, consider the eigenvalue decomposition of $\rho$
\begin{equation}
\rho = \sum_{i=0}^{2^Q-1} p_i |\psi_i\rangle\langle\psi_i|,
 \label{eq:eigen}
\end{equation}
with $p_i$ decreasing with increasing $i$, and $Q$ being the number of the qubits. Then,
\begin{equation}
\hat{\mu}^{\rm VD} = \frac{\langle \psi_0 | X | \psi_0 \rangle+\sum_{i=1}^{2^Q-1} (p_i/p_0)^M\langle \psi_i | X | \psi_i \rangle}{1+\sum_{i=0}^{2^Q-1} (p_i/p_0)^M}.
\end{equation}
Therefore, we can see that  VD exponentially suppresses contributions to $\hat{\mu}^{\rm VD}$ from eigenvectors of $\rho$ other than the dominant one $|\psi_0\rangle$. While increasing $M$ suppresses the errors, for large error rates it also exponentially increases the shot cost of the estimation of $\hat{\mu}^{\rm VD}$  with a given accuracy~\cite{czarnik2021qubit}.

VD implicitly assumes that the dominant eigenvector $|\psi_0\rangle$ is close to the noiseless state $|\psi_{\rm exact}\rangle=U\ket{0}^{\otimes Q}$. 
Therefore, the quality of the VD correction is limited by a noise floor
\begin{equation}
\epsilon = \langle \psi_0 | X | \psi_0 \rangle - \langle \psi_{\rm exact} | X | \psi_{\rm exact} \rangle.
\label{eq:noise_floor}
\end{equation}
The noise floor can be bounded from above using a quantity called the coherent mismatch
\begin{equation}
c = 1 - |\langle \psi_{\rm exact} | \psi_0 \rangle |^2,  
\end{equation}
i.e.,
\begin{equation}
|\epsilon| \le 2\sqrt{c} ||X||_{\infty}, 
\label{eq:VD_bound}
\end{equation}
where $||X||_{\infty}$ is the absolute value of  the largest eigenvalue of $X$~\cite{koczor2021dominant}.  When $|\psi_0\rangle$ is an eigenvector of $X$, the  bound (\ref{eq:VD_bound}) is replaced~\cite{koczor2021dominant} by
\begin{equation}
|\epsilon| \le 2c ||X||_{\infty}
\label{eq:VD_bound_MC}
\end{equation}
leading to better-than-expected performance of VD.
For realistic noise, $\epsilon$ is usually larger than $0$~\cite{koczor2020exponential}. Nevertheless, it was shown that for sufficiently small  error rates   the noise floor does not prevent VD from significantly reducing the impact of noise~\cite{koczor2021dominant}.

Here we recall that ${\rm Tr }[\rho^M X]$ can be computed using $M$ copies of the noisy state $\rho$ and an ancillary qubit with a controlled permutation of the copies which changes the position of each copy (so-called controlled derangement). Consequently, $QM+1$ qubits are required to mitigate a $Q$-qubit state.  This number  can be reduced to $2Q+1$ by employing qubit resets and increasing the depth of the circuit performing the controlled derangement~\cite{czarnik2021qubit}. The controlled derangement  can be implemented as a circuit of $QM$ controlled swaps. Therefore,  the derangement introduces errors to the estimates of ${\rm Tr }[\rho^M X]$ and ${\rm Tr }[\rho^M]$  which may require correction with other error mitigation methods~\cite{koczor2020exponential}.

Alternative approaches utilizing state verification to implement VD while further reducing required qubit counts were proposed recently~\cite{huo2021dual, cai2021resourceefficient}. Furthermore, approaches   conceptually unifying VD with both ZNE and subspace expansion techniques were proposed ~\cite{xiong2021quantum, yoshioka2021generalized, cai2021quantum, mari2021extending}.

\section{Unifying variable noise Clifford Data Regression methods with Virtual Distillation}

As we have discussed in the previous section, results obtained with CDR and vnCDR show that it is beneficial to train an error mitigation scheme  with training data obtained from near-Clifford circuits, as well as from data arising from running  circuits under different  noise levels. Here, we propose to additionally use data from VD with multiple numbers of copies.  This conceptually unifies VD, ZNE, and CDR into one mitigation method which we call UNIfied Technique for Error mitigation with Data (UNITED).

Let us now explain how the dataset for UNITED is generated. Given a set of near-Clifford circuits similar to the circuit of interest $U$, $\{V_i\}_{i=1}^{N_t}$, we first use a classical computer to calculate the noiseless expectation values (denoted by $\mu_i$) of the observable of interest $X$ for each circuit. Then, we estimate the  noisy and VD mitigated expectation values of $X$ for different  noise levels $c_j$ with $j=0,\ldots, n$ (as in ZNE), and for different numbers of copies $M=1,\ldots,M_{\rm max}$. These estimates are made using a noisy quantum computer. Note that $M=1$ labels the standard noisy estimates as they are obtained using a single copy.  Putting all this together, we obtain a training set $\mathcal{T}^{\rm UNITED} = \{(\hat{\vec{\mu}}^{\rm UNITED}_i, \mu_i)\}_{i=1}^{N_t}$, where $\hat{\vec{\mu}}^{\rm  UNITED}_i = (\hat{\mu}_{i,0,1},\dots,\hat{\mu}_{i,n,M_{\rm max}})$ is a tensor of the noisy and  VD mitigated estimates for $V_i$, and  $\hat{\mu}_{i,j,M}$ is the expectation value obtained at  noise level $c_j$ and with $M$ copies. We use $\mathcal{T}^{\rm UNITED}$ to train a linear function $f^{\rm UNITED}: (\mathcal{E}_{\rm VD})^{ M_{\rm max} (n+1)}\rightarrow\mathcal{E}_{\rm noiseless}$ of the form 
\begin{equation}
       f^{\rm UNITED}(\vec{x}) = \sum_{j=0}^{n}\sum_{M=1}^{M_{\rm max}} d_{j,M} x_{j,M}\,.
\label{eq:UNITED_ansatz}
\end{equation} 
Here we use $\mathcal{E}_{\rm VD}$ to denote a space of the VD-mitigated expectation values. For simplicity of notation, $\mathcal{E}_{\rm VD}$ also includes the noisy expectation values ($M=1$). 

\begin{table}[t]
\small
\centering
\begin{tabular}{|l|ccc|}
\hline
\textbf{Technique}           & \multicolumn{1}{c|}{\textbf{Noise Levels}} & \multicolumn{1}{c|}{\textbf{Copies}} & \textbf{Training} \\ 
         & \multicolumn{1}{c|}{} & \multicolumn{1}{c|}{} & \textbf{Circuits} \\ \hline
ZNE~\cite{temme2017error}     & \multicolumn{1}{c|}{\cmark}          & \multicolumn{1}{c|}{\xmark}     & \xmark                \\ \hline
VD~\cite{koczor2020exponential,huggins2020virtual}       & \multicolumn{1}{c|}{\xmark}           & \multicolumn{1}{c|}{\cmark}    & \xmark                \\ \hline
CDR~\cite{czarnik2020error}       & \multicolumn{1}{c|}{\xmark}           & \multicolumn{1}{c|}{\xmark}     & \cmark               \\ \hline
vnCDR~\cite{lowe2020unified}    & \multicolumn{1}{c|}{\cmark}          & \multicolumn{1}{c|}{\xmark}     & \cmark               \\ \hline
CGVD [Here]     & \multicolumn{1}{c|}{\xmark}           & \multicolumn{1}{c|}{\cmark}    & \cmark               \\ \hline
UNITED [Here]   & \multicolumn{1}{c|}{\cmark}          & \multicolumn{1}{c|}{\cmark}    & \cmark               \\ \hline
\end{tabular}
\caption{\textbf{Resource requirements for the different error mitigation strategies explored in this work.} In this table we show the necessary ingredients to employ the different error mitigation strategies that we unify and benchmark in this work.}
\label{tab:diff}
\end{table}

We use least-squares regression  to find the values $d^{*}_{j,M}$ that correspond to the best possible fit of $\mathcal{T}^{\rm UNITED}$. This fitted ansatz is then used to obtain  an error-mitigated expectation value of $X$ for the circuit of interest as 
$\mu^{\rm UNITED}=f^{\rm UNITED}(\vec{\hat{\mu}}^{\rm UNITED})$. Here,  $\hat{\vec{\mu}}^{\rm UNITED} = (\hat{\mu}_{0,1},\dots,\hat{\mu}_{n,M_{\rm max}})$ is a tensor of the VD-mitigated and the noisy expectation values of $X$ for $U$, where $\hat{\mu}_{j,M}$ is obtained at the noise level $c_j$ with $M$ copies.  It is not hard to see that under this framework, vnCDR is a special case of  UNITED. Indeed, for the special case of $M_{\rm max}=1$ $\mathcal{T}^{\rm UNITED}$ becomes the  vnCDR dataset  $\mathcal{T}^{\rm vnCDR}$. For the reader's benefit, we summarize the similarities and differences between UNITED and other error mitigation methods in Table~\ref{tab:diff}.

While we have presented the UNITED method, we still need to show that it is well-motivated. In what follows we will study special cases of UNTIED that illustrate how its different moving parts help to mitigate errors.

\subsection{Error mitigation from VD data with multiple copy numbers}
\label{sec:CGVD_ansatz}

 In this section, we consider a special case of the ansatz used in the UNITED method where we do not use multiple noise levels ($n=0$). That is, we show that one can use a combination of  the VD mitigated observables with different numbers of copies as an estimate of the noiseless observable that has an error smaller than each VD mitigated observable. For this purpose, we consider a toy  noise model where for each gate an error occurs with probability $\lambda$, and every different sequence of errors results in an orthogonal state.
Additionally, at the end of the circuit we place a coherent error given by a unitary $U(\theta) = e^{-i \theta E}$, where $E$ is a Hermitian operator. Such an error produces a non-zero noise floor for $\theta \neq 0$, while preserving the orthogonality of states corresponding to different sequences of the gate errors. 
We describe this noise model in more detail in Appendix~\ref{app:VD_comb}.
Furthermore, in the following, we neglect the errors caused by  an imperfect derangement operation and the shot noise.

First, let's consider taking a linear combination of VD-mitigated expectation values of $X$ for $U$  obtained using various $M$ up to $M_{\rm max}$. We use such a linear combination to construct an estimate of $\mu$ that  improves on  all the VD mitigated estimates. That is, we take 
\begin{equation}
\mu'_{M_{\rm max}} = \sum\limits_{M=1}^{M_{\rm max}}b_{M}\hat\mu_{M}^{\rm VD} \ ,
\label{eq:comb_VD0}
\end{equation}
where $b_{M}$ are some coefficients to be found.
 In Appendix~\ref{app:VD_deriv} we show that for this noise model the orthogonal error scales as  
\begin{equation}
   | \mu_{M}^{\rm VD}  - \mu - \epsilon | \in \mathcal{O}(\lambda^{M}) \ ,
\end{equation}
while  one can find coefficients $b_M$ (see  Appendix~\ref{app:VD_comb}) which give  
\begin{equation} \label{eq:comb-VD}
  |  \mu_{M_{\rm max}}' - \mu - \epsilon | \in \mathcal{O}(\lambda^{M_{\rm max}+1})\,.
\end{equation}
Therefore,  $\mu'_{M_{\rm max}}$ can be used to suppress the orthogonal error beyond the error of the VD estimate with the largest number of copies.   We note that the ansatz used here is similar to the one used by the Richardson extrapolation~\eqref{Richardson} with higher noise levels $c>1$ replaced by larger copy numbers $M>1$.

\subsection{Error mitigation from  VD data with multiple copy numbers guided by near-Clifford circuits}

Here we consider a special case of UNITED which corresponds to taking $n=0$ and $M_{\rm max}>1$. This is equivalent to fitting the ansatz considered above (\ref{eq:comb_VD0}) using a training set obtained from near-Clifford circuits. 
First, we start by noting that even in the case of the simple noise model considered in the previous section, one needs detailed knowledge of the  noise model  to determine the coefficients $b_{M}$ (see Appendix~\ref{app:VD_comb}). However, such knowledge would require expensive process tomography in the case of a real device.   Therefore, we propose to use data from near-Clifford quantum circuits to fit the coefficients $b_{M}$ that lead to the desired orthogonal error suppression. We note that this corresponds to  constructing a
training set  analogously to the case of vnCDR but replacing the $c>1$ noisy expectation values with VD-mitigated ones. Furthermore, the same as in the case of vnCDR, we use a linear ansatz. We call this approach Clifford Guided Virtual Distillation (CGVD). 

In more detail, the CGVD training set for an observable of interest $X$ and a circuit of interest $U$ is defined as $\mathcal{T}^{\rm CGVD} = \{(\hat{\vec{\mu}}_i^{\rm CGVD}, \mu_i)\}_{i=1}^{N_t}$, where $\hat{\vec{\mu}}_i^{\rm CGVD} = (\hat{\mu}_{i,1}^{\rm CGVD},\dots,\hat{\mu}_{i,M_{\rm max}}^{\rm CGVD})$ is a vector of the  VD-mitigated and noisy  expectation values of $X$  for a near-Clifford circuit $V_i$, which  are obtained at the base noise level $c=1$ and with $M=1,\dots,M_{\rm max}$. The training set is fitted with a linear function $f^{\rm CGVD}:(\mathcal{E}_{\rm VD})^{M_{\rm max}}\rightarrow\mathcal{E}_{\rm noiseless}$,  
of the form 
\begin{equation}
       f^{\rm CGVD}(\vec{x}) = \sum_{M=1}^{M_{\rm max}} b_M x_{M}\,,
\label{eq:CGVD_ansatz}
\end{equation}
where the parameters $b_{M}$ are found using least-squares regression. This leads to the best-fit parameters $b^{*}_{M}$ which  are then used to correct the noisy and VD-mitigated estimates of $X$ expectation value for the circuit of interest, i.e. $\mu^{\rm CGVD} = f^{\rm CDVD}((\hat\mu, \dots, \hat \mu^{VD}_{M_{\rm max}}))$. 
 We note that CGVD perfectly mitigates global depolarizing noise as shown in Appendix~\ref{app:CGVD_global}.  Such  perfect correction  is possible as for this noise model the noise floor is zero.  

\subsection{UNITED ansatz for a simple noise model with orthogonal and coherent  errors}

In this section we demonstrate  that by combining ZNE, VD and CDR, the UNITED method can suppress the error better than  its  individual parts.  For this purpose, we consider again a simple noise model from Section~\ref{sec:CGVD_ansatz}. 
Let us construct for this model  a dataset  $\vec{\hat \mu}^{\rm UNITED}$, which is created  using  $n+1$  noise levels and $M_{\rm max}$ copy numbers. For a noise level $c_j$ we scale the noise strengths   $\lambda$ and $\theta$  as  $\lambda_j = c_j \lambda_0$ and $\theta_j = c_j \theta_0$, where $\lambda_0$ and $\theta_0$ are the base noise strengths. As described in detail in Appendix~\ref{app:VD_noise_comb}, for each one of these noise levels  we suppress the orthogonal gate errors with  a combination of   VD mitigated observables of form~\eqref{eq:comb_VD0}, i.e.
\begin{equation}
\mu'_{M_{\rm max},j} = \sum_{M=1}^{M_{\rm max}} b_M \hat \mu_{j,M}\,.
\end{equation}
From Eq.~\eqref{eq:comb-VD}  we obtain
\begin{equation}
  |  \mu'_{M_{\rm max},j}  - \mu -\epsilon_{j}| \in \mathcal{O}(\lambda_{j}^{M_{\rm max}+1})\,,
\end{equation} 
where $\epsilon_{j}$ is the noise floor for the noise level $c_j$. As shown in Appendix~\ref{app:VD_noise_comb}, for such obtained estimates $\{\mu'_{M_{\rm max},j}\}_{j=0}^n$ one can perform a Richardson extrapolation to additionally suppress the coherent error resulting in the noise floor.

By  doing that one gets  the final estimate of the expectation value of interest 
\begin{equation}
\mu''_{M_{\rm max},n} = \sum_{j=0}^{n} \gamma_j \mu'_{M_{\rm max},j} = \sum_{j=0}^{n} \sum_{M=1}^{M_{\rm max}} \gamma_j b_M
\hat \mu_{j,M},
\end{equation}
where $\gamma_j$ coefficients are given by Eq.~\eqref{eq:Richardson_coeff}, that for large enough $M_{\rm max}$  has an error 
\begin{equation}
|\mu''_{M_{\rm max},n} - \mu| \in \mathcal{O}(\theta_{0}^{n+1}).
\end{equation}
 For a more detailed explanation of  the final error see Appendix~\ref{app:VD_noise_comb}. 
This improved estimate has a form of the UNITED ansatz~\eqref{eq:UNITED_ansatz} motivating its choice. We note that,  similarly to the case of CGVD, even in the case of our simple noise models determining the ansatz coefficients requires detailed knowledge of the noise model that is usually unavailable for real-word quantum computers. Therefore, in the UNITED methods we aim to find these coefficients by learning from near-Clifford training circuits.    

\subsection{General Remarks}

It is important to note that although a combination of noisy observables inspired by Richardson extrapolation motivates the form of the  ansatz in Eq.~\eqref{eq:UNITED_ansatz}, it does not guarantee good performance of the new method for several reasons. Firstly, all methods to scale physical noise are only approximate so the noise levels will be imperfectly controlled, leading to imperfect extrapolated estimates. Nevertheless, successful applications of ZNE and vnCDR show that imperfect noise scaling control does not preclude successful error mitigation. Very importantly, training a model based on near-Clifford circuit data is not equivalent to performing the Richardson extrapolation, so the performance of UNITED depends on the ability to construct informative training sets. 
However, analogous to the case of vnCDR, we expect that fitting ansatz coefficients~\eqref{eq:UNITED_ansatz} instead of performing an extrapolation is beneficial as it does not require knowledge of noise levels, reducing the impact of imperfect noise control.  

Overall, combining observables with different  noise levels can be used to construct estimates with less error as shown by Richardson extrapolation. However, the current level of noise in quantum computers is significant. Due to the concentration of the expectation values of observables for high noise levels, a large number of shots is necessary to extract  information from highly noisy data~\cite{wang2020noise,franca2020limitations}.  
This limits the range of the noise levels that can be used in practice. Therefore, at an intuitive level, using VD-mitigated observables is another way to provide data corresponding to  different noise levels enhancing previous   error mitigation methods using variable noise levels, like ZNE and vnCDR. From this perspective, the UNITED ansatz uses near-Clifford circuits to learn the best way to combine the various observables calculated under different noise levels. 

Another important remark is that adding copy data to the training set further increases the shot cost of the mitigation. Therefore, the best choice of noise levels and the optimal number of copies is not known a priori. Despite this, we found that the method gives mitigated estimates, which  remain stable with increasing $n$ and $M_{\rm max}$ and that it either outperforms or matches vnCDR performance for sufficiently large shot numbers. 

\section{Numerical Results} 
In this section we present the performance of ZNE, vnCDR, VD, and UNITED applied to noisy outputs obtained from simulations of a trapped-ion quantum computer with a realistic noise model.  We consider the task of mitigating the output of random quantum circuits and the output of the Quantum Alternating Operator Ansatz applied to Max-Cut problems. We use these two settings to benchmark the mitigation methods for various system sizes and different circuit depths. We study several total shot budgets per mitigated expectation value, highlighting the strengths and weaknesses of each method.

\subsection{Random quantum circuits}

\begin{figure}[t]
\includegraphics[width=\columnwidth]{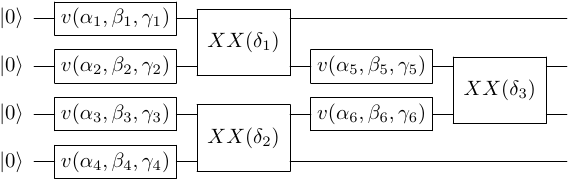}
\caption{\textbf{Random quantum circuit structure used for our  benchmark simulations.}  Here we show a layer of an exemplary  random quantum circuit for $Q=4$. The layer is built from two layers of alternating nearest-neighbor Molmer-Sørensøn gates $XX(\delta)=e^{-i\delta \sigma_X^j \sigma_X^{j+1}}$ gates, where $\sigma_X^j$ is a Pauli operator acting on qubit~$j$. The $XX$ gates are decorated with general single-qubit unitaries  $v(\alpha,\beta,\gamma) = R_Z(\alpha) R_Y(\beta) R_Z(\gamma)$. Here $R_Z(\alpha) = e^{-i\alpha/2 \sigma_Z}$, $R_Y(\beta) = e^{-i\beta/2 \sigma_Y}$, and $\sigma_Z,\sigma_Y$ are Pauli operators. We  choose angles $\alpha,\beta,\gamma,\delta$ randomly for each random quantum circuit instance of the gates.  We remark that $R_Y,R_Z,XX$ are native gates of a trapped-ion quantum computer. } 
\label{fig:RQC}
\end{figure}

We investigate random quantum circuits for various numbers of qubits $Q=4,6,8,10$ and several circuit depths. We consider random quantum circuits constructed using the native gates of the trapped-ion quantum computer we are modeling, namely of $L$ layers of alternating nearest-neighbor Molmer-Sørensøn entangling gates decorated with general single qubit unitaries, see Figure~\ref{fig:RQC} for details. Both the entangling gates and the single qubit unitaries are parametrized by randomly chosen angles. Here we simulate random quantum circuits with a number of layers $L=gQ$, taking $g=1$ and $g=16$.

We expect that for this setup, random quantum circuits will output states that converge to random states sampled according to Haar measure~\cite{mcclean2018barren}, with increasing $L$. Therefore, the setup is general and challenging, making it relevant for testing error mitigation methods. Furthermore, it is worth noting that deep circuits implemented in a trapped-ion quantum computer present a favorable setting for VD as this architecture enables all-to-all connectivity which decreases the depth of the controlled derangement circuit compiled to native gates~\cite{maslov2017basic}. Furthermore, it is expected that for sufficiently deep circuits and sufficiently small gate error rates, the coherent mismatch does not prevent a good quality mitigation with VD~\cite{koczor2021dominant}.

\begin{figure*}[t]
\centering
\includegraphics[width=0.9\textwidth]{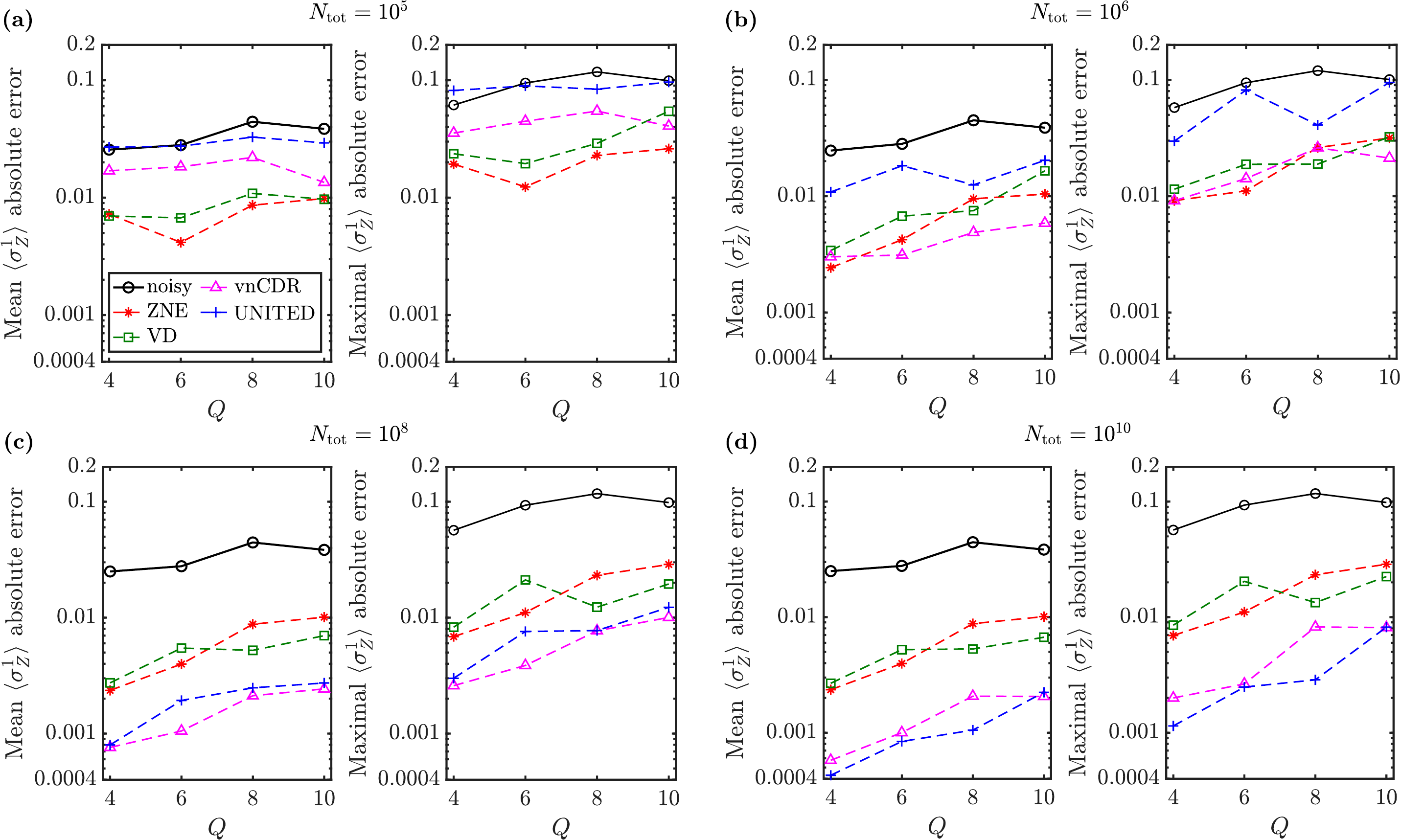}
\caption{\textbf{Comparing error mitigation methods applied to RQCs for a number of layers $L=Q$.} We show the mean and maximal absolute errors of the expectation value of $\sigma_Z$ measured at the first qubit of $30$ RQC instances for the noisy estimates and estimates obtained with different error mitigation methods. In (a), the results for  the total number of shots per mitigated (noisy) expectation value $N_{ \rm tot}=10^5$. In (b), $N_{ \rm tot}=10^6$. In (c), $N_{ \rm tot}=10^8$. In (d), $N_{ \rm tot}=10^{10}$.  For $N_{ \rm tot}=10^5$, we find that ZNE gives the best results.  For $N_{ \rm tot}=10^6$, $10^8$ vnCDR gives the best results while  for $N_{ \rm tot}=10^{10}$ UNITED gives the best results.
} 
\label{fig:g1}
\end{figure*}

\begin{figure}[t]
\includegraphics[width=\columnwidth]{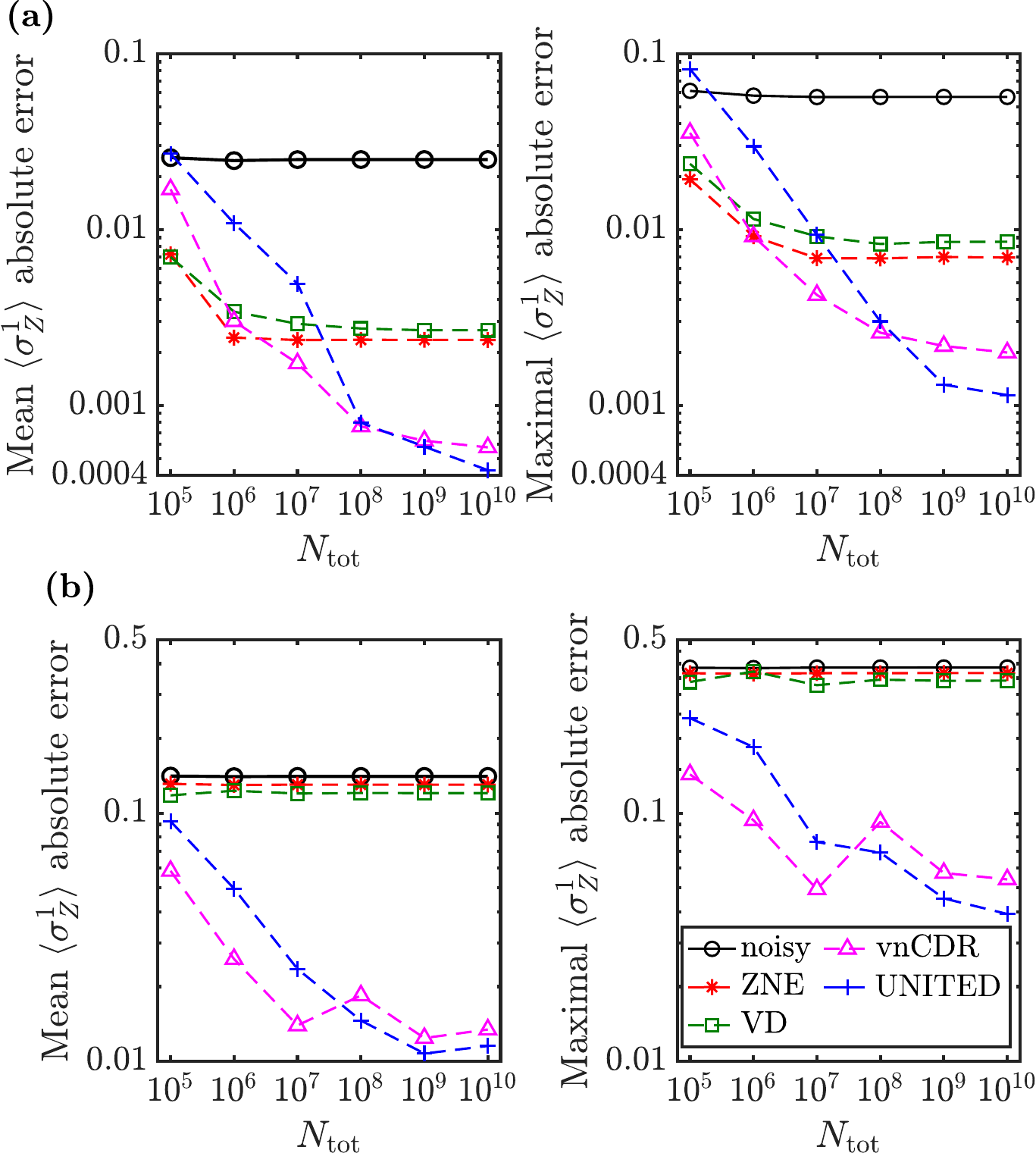}
\caption{\textbf{Convergence of error mitigation methods applied to RQCs with increasing $N_{\rm tot}$.}  Mean and maximal absolute  errors of the expectation value of  $\langle \sigma_Z^1 \rangle$  for random quantum circuits plotted versus the shot budget.  We use $30$ instances of random quantum circuits. In (a) we show results for  $Q=4, g=1$. In (b) we display the results for $Q=4, g=16$, where the number of layers $L=gQ$. For $Q=4, L=4$ we find that ZNE and VD estimates converge first. The estimates produced by vnCDR converge slower but outperform those given by VD and ZNE for shot budgets of $10^{7}-10^{10}$. The convergence of UNITED is the slowest but it gives the best results for $N_{\rm tot}=10^{9},10^{10}$. For $Q=4, L=64$ estimates produced by ZNE and VD basically offer no mitigation and are outperformed by vnCDR and UNITED for all considered shot budgets. We see an improvement of UNITED over vnCDR for $N_{\rm tot}=10^{9},10^{10}$. A further discussion of the performance of VD for different shot budgets is explored in Appendix \ref{app:VD_copy_analysis}. 
} 
\label{fig:conv}
\end{figure}

 \subsection{Random quantum circuit results }\label{sec:RQC result}
 
First, we analyze the case of $L=Q$ ($g=1)$. For this choice, we consider system sizes of $Q=4,6,8,10$. For each pair of $L$ and $Q$ values we consider $30$ instances of random quantum circuits and for each random quantum circuit, we mitigate the expectation value of $\sigma_Z$ acting on the first qubit of the circuit $\langle \sigma_Z^1 \rangle$. We generate noisy data using a realistic trapped-ion  computer noise model described in Appendix~\ref{app:noise_Model}. In general, deep RQCs will produce observables that are clustered around $0$ and noise makes this clustering phenomenon even stronger \cite{mcclean2018barren, wang2020noise}. However, the  RQCs we explore here are not sufficiently deep to observe this feature, see Appendix~\ref{app:clustering} for details.  

We perform the mitigation with data obtained using a finite number of shots. For each mitigated expectation value and each error mitigation method, we assign the total number of shots used by the method to mitigate the expectation value. This number is defined as the total shot budget $N_{\rm tot}$. We consider $N_{\rm tot} = 10^5-10^{10}$. 

Different error mitigation methods require estimates of observables from a different total number of circuits. In the case of ZNE, the circuit of interest needs to be evaluated for several different state preparation noise levels. In the case of vnCDR, estimates for the observable of interest are required for many training circuits evaluated at several noise levels. UNITED additionally requires performing VD mitigation for both the training circuits and the circuit of interest. Therefore, for the same shot budget, different mitigation methods have different numbers of shots per circuit used in the method. Consequently, for a given total number of shots, the accuracy of individual observable estimates decreases as the required total number of circuits for a given error mitigation method increases. We note that for fixed $N_{\rm tot}$ we can distribute shots between different observable evaluations in many ways.  Here for the sake of simplicity for a given method,  we assign the number of shots to be uniformly distributed across all the required circuits. That is the number of shots per circuit to run for a given mitigation method is the total number of shots divided by the total number of circuits necessary to run.  

In Figure~\ref{fig:g1} we plot the mean and maximal absolute errors of $\langle \sigma_Z^1 \rangle$. We find that the performance of the error mitigation methods depends strongly on the shot budget. The mitigated observables produced by ZNE and VD converge fastest with increasing $N_{ \rm tot}$. Furthermore, ZNE offers the best correction for the smallest total shot budget $N_{ \rm tot}=10^5$. The convergence of vnCDR is slower but it outperforms ZNE and the other methods for $N_{ \rm tot}=10^6$  up to $N_{ \rm tot}=10^8$. We find that UNITED exhibits the slowest convergence with increasing $N_{ \rm tot}$ but it gives the best results for $N_{ \rm tot}=10^{10}$. The quality of the mitigated results decreases with increasing $Q$ as the number of noisy gates increases. Nevertheless, even for the largest considered $Q=10$, a factor of $20$ improvement over the noisy result can be obtained with vnCDR and UNITED. The convergence of the results with $N_{\rm tot}$ is more explicitly shown for $g=1$, $Q=4$ in Figure~\ref{fig:conv}(a).

Next, we analyze the case of $L=16Q$ ($g=16)$. For this choice, we simulate  $Q=4,6,8$. Again, for each pair of $L$ and $Q$ values, we consider $30$ instances of random quantum circuits and mitigate $\langle \sigma_Z^1 \rangle$ for each of them. In this regime, the appearance of clustering of the exact observables, due to the properties of RQCs, becomes apparent in the largest system size, see Appendix~\ref{app:clustering} for details.

As in the case of $g=1$, we perform the mitigation for $N_{\rm tot} = 10^5-10^{10}$.
For $Q=6, L = 96$ and $Q=8, L = 128$ we were unable to obtain improvement over the noisy results with any error mitigation method as the number of noisy gates is too large.
For smaller systems of $Q=4, L=64$  we find that learning-based error mitigation methods outperform VD and ZNE  even for  $N_{\rm tot} = 10^5$. For $N_{\rm tot} = 10^7-10^{10}$ we find that both vnCDR and UNITED give a factor of $5-10$ improvement of the error in the noisy results. For $N_{\rm tot} = 10^9-10^{10}$  we see an improvement of UNITED over vnCDR. Again we find that ZNE and VD estimates converge fastest with increasing  $N_{\rm tot}$. Again, the convergence of UNTIED is the slowest but it also provides the best mitigation quality for the largest shot budget. We show $g=16$, $Q=4$ convergence results in detail in Figure~\ref{fig:conv}(b). 

For this high noise regime, learning-based error mitigation is particularly sensitive to shot noise \cite{czarnik2022improving} as noisy expectation values in the training set become more clustered with increasing noise strength. Different runs with different shot noise instances can lead to large differences in the mitigated observable. Therefore, one can expect that the behavior of the mean and maximal errors of the mitigated expectation values for a sample of RQCs are very sensitive to finite sample size effects. Consequently, the peak around $N_{\rm tot} = 10^8$ observed in the convergence of vnCDR  with an increasing number of shots in~Figure~\ref{fig:conv}(b) is consistent with the expected behavior in this regime. 

\begin{figure*}
    \centering
    \includegraphics[width=0.9\textwidth]{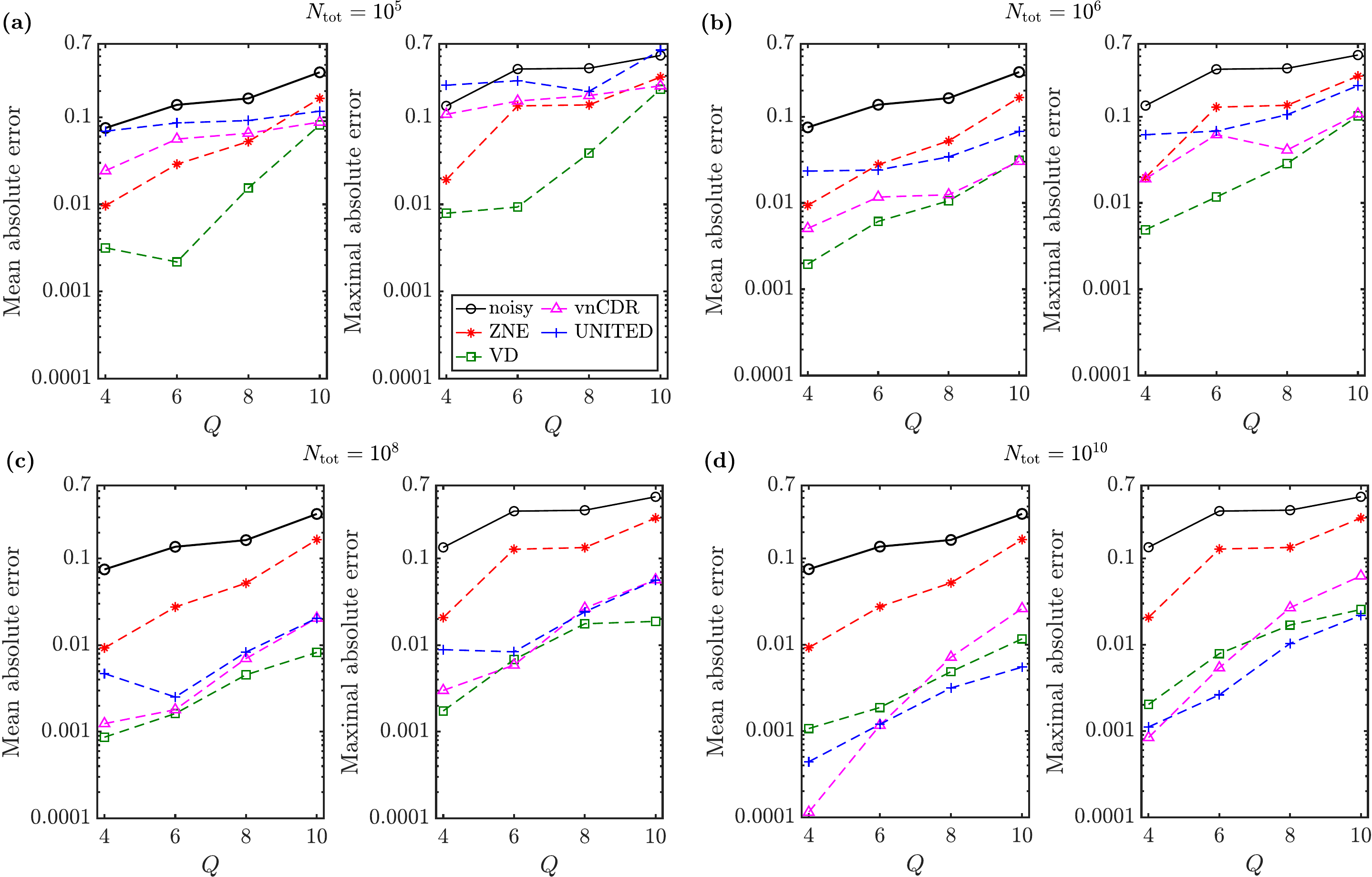}
    \caption{\textbf{Comparing error mitigation methods applied to Max-Cut QAOA.} Mean and maximal absolute errors of the expectation value of a randomly selected term of the Max-Cut Hamiltonian on 28 instances of Erd\"os-Renyi graphs. Here we compare the noisy estimates and the error-mitigated ones obtained with various error mitigation methods. In (a), the results for the total number of shots per mitigated (noisy) expectation value at $N_{\textrm{tot}}=10^5$. In (b) $N_{\textrm{tot}}=10^6$, (c) $N_{\textrm{tot}}=10^8$ and in (d) $N_{\textrm{tot}}=10^{10}$. For $N_{\textrm{tot}}=10^5,\ 10^6,\ 10^8$, VD gives the best result while for $N_{\textrm{tot}}=10^{10}$ UNITED gives the best results. } 
    \label{fig:MC_qubits}
\end{figure*}

\subsection{ QAOA for Max-Cut problems}
In this section, we investigate mitigating observables arising from the application of the Quantum Alternating Operator Ansatz (QAOA)~\cite{farhi2014quantum, hadfield2019quantum} to solve Max-Cut problems. We consider a graph $G = (V,E)$ with vertices in $V$, $|V|=Q$, and edges in $E$.  The  Max-Cut problem is to find   a bipartition of the graph that maximizes  the number of edges connecting the halves.  It can be solved by minimizing the energy of a Hamiltonian
\begin{equation}
H_P = \sum_{(i,j) \in E} \sigma_Z^i \sigma_Z^j,  
\label{eq:maxcut}
\end{equation}
where we are summing over the edges.  Here we consider Erd\"os-Renyi graphs~\cite{erdos1959random} which connect each pair of nodes $(i,j)$ with probability $p_e$.
The QAOA ansatz is built with $L$ layers 
\begin{equation} \label{eq:qaoa}
\begin{split}
|\Psi(\vec{\alpha},\vec{\beta})\rangle &= e^{-i \beta_L H_M} e^{-i \alpha_L H_P} \dots \\
&\dots e^{-i \beta_2 H_M} e^{-i \alpha_2 H_P} e^{-i \beta_1 H_M} e^{-i \alpha_1 H_P} |\vec{+} \rangle
\end{split}
\end{equation}
where $\vec{\alpha}= (\alpha_1, \alpha_2, \dots, \alpha_L)$,    $\vec{\beta}= (\beta_1, \beta_2, \dots, \beta_L)$ are parameters of the ansatz,   
$H_M = \sum_{i=1}^Q \sigma_X^i$, and $|\vec{+} \rangle$ is a tensor product state of single-qubit $|+\rangle = \frac{1}{\sqrt{2}} ( |0\rangle + |1\rangle)$ states.

Here we consider $Q=4,6,8,10$ and  $L=Q$. For each $Q$ we construct $28$ 
Erd\"os-Renyi graphs, taking $p_e=0.5$, obtained by the uniform sampling of their distribution.  For each graph, we perform  noiseless optimization of the Max-Cut cost function $\langle H_P \rangle$ to find optimal $(\vec{\alpha}_0,\vec{\beta}_0)$.  Then, we mitigate the expectation value $\bra{\psi(\vec{\alpha}_0,\vec{\beta}_0)}\sigma_Z^i \sigma_Z^j \ket{\psi(\vec{\alpha}_0,\vec{\beta}_0)}$ for one randomly selected $(i,j)\in E$. We compile $\ket{\psi(\vec{\alpha}_0,\vec{\beta}_0)}$ to the trapped-ion native gate set and obtain the noisy expectation value using the trapped-ion noisy simulator. Here we  assume full device connectivity. Additional details on the compilation are given in Appendix~\ref{app:comp}. For $Q=10$ the compiled circuits  have  on average $96$ layers of gates and $227$ 2-qubit  Molmer-Sørensøn  gates. We note that since the number of edges in Erd\"os-Renyi graphs is a random variable, the structure of the compiled circuits varies from instance to instance. Hence, we give average values of the depth and the gate count here.

\subsection{QAOA results}
\label{sec:qaoa}
We mitigate the expectation value of a random term of $H_P$ for each instance of Erd\"os-Renyi graph. Once again, we employ the trapped-ion noise model described in Appendix~\ref{app:noise_Model}. In Figure~\ref{fig:MC_qubits} we show the mean and maximal absolute errors of the 
mitigated observables over the $28$ graph instances. We again find (as in the case of random quantum circuits) that the performance of the error mitigation methods depends primarily on the shot budget. Interestingly, here we find that VD performs much better than in the case of random quantum circuits and we observe that it is the best-performing method for $N_{\textrm{tot}}=10^5,10^6,10^8$ shot budgets. ZNE is the second best method for $N_{\textrm{tot}}=10^5$ but fails to improve significantly for larger shot budgets. The learning-based methods improve systematically with increasing $N_{\rm tot}$. For the largest considered $N_{\rm tot} = 10^{10}$, UNITED is the best method for $Q=6,8,10$ and vnCDR is the best for $Q=4$. 

\begin{figure}
    \centering
    \includegraphics[width=\columnwidth]{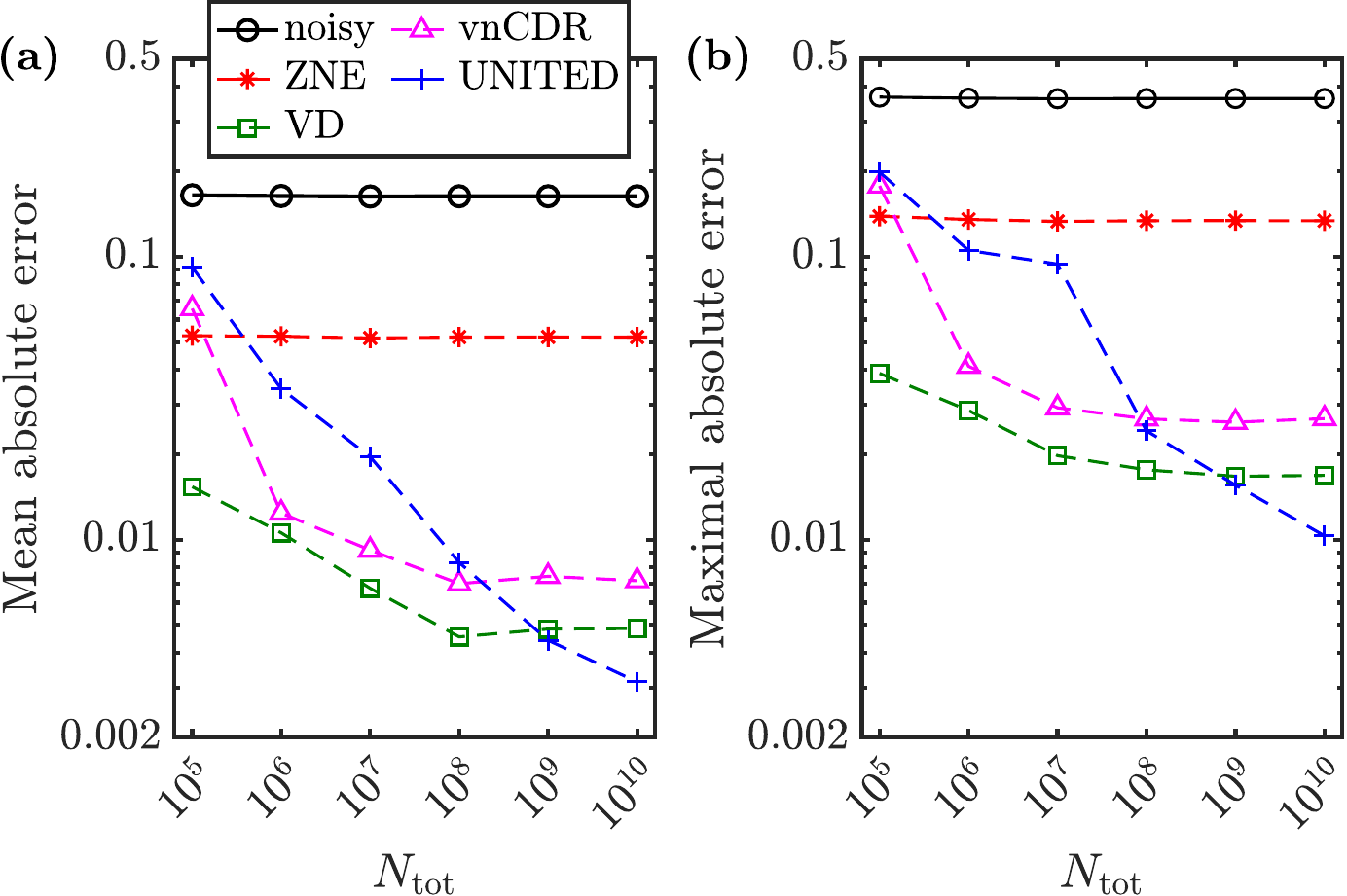}
    \caption{\textbf{Convergence of error mitigation methods applied to the Max-Cut QAOA for $Q=8$ with increasing $N_{\textrm{tot}}$.} Mean (a) and maximal (b) absolute errors of the expectation value of a random term of the Max-Cut Hamiltonian on 28 instances of Erd\"os-Renyi graphs. ZNE has already converged at $N_{\textrm{tot}}=10^5$, while VD is the best performer up to its convergence at $N_{\textrm{tot}}=10^8$, after which $(N_{\textrm{tot}}=10^9)$ UNITED becomes the best performer. vnCDR converges at $N_{\textrm{tot}}=10^8$.}
    \label{fig:MC_budget}
\end{figure}

Figure~\ref{fig:MC_budget} analyzes the convergence of the various techniques for $Q=8$ in more detail. Here VD is the best performer at the smallest shot budget ($N_{\textrm{tot}}=10^5$), followed by ZNE, vnCDR, and UNITED. ZNE shows no improvement with an increased shot budget. UNITED, vnCDR, and VD, on the other hand, continue improving with UNITED becoming better than vnCDR at $N_{\textrm{tot}}=10^8$ and vnCDR seeing no further improvement after this point. Similarly, VD sees continued improvements up to $N_{\textrm{tot}}=10^8$, being the best technique up to that point. From $N_{\textrm{tot}}=10^9$ and above, UNITED is the best method. 

The reason for the better performance of VD when compared to the random quantum circuit case is likely due to the fact that the solutions to the Max-Cut problems are eigenstates of the mitigated observables. Therefore, we expect the noise floor to be reduced relative to the random quantum circuits (see (\ref{eq:VD_bound}) and (\ref{eq:VD_bound_MC})). 

We note that the relative performance of VD with respect to vnCDR depends on the application. In particular, VD performance is better for QAOA than for the RQC experiments. This coincides with a larger improvement of UNITED over vnCDR for QAOA. Therefore, these two numerical studies highlight the utility of unifying error mitigation methods. Leveraging the strengths of ZNE, CDR, and VD into one method appears to allow for better performance across a range of tasks.

\section{Implementation Details}
\label{sec:impl}

\subsection{Zero Noise Extrapolation}

In the simulations described above, when implementing ZNE we use two noise levels $c \in \{1,2\}$ and linear extrapolation. For the case of random quantum circuits, we compared this approach  to 
exponential extrapolation (defined as in Ref.~\cite{giurgica2020digital}) and linear extrapolation with $c \in \{1,2,3\}$  and found that the method we employ works best in general. We present  details of this comparison in  Appendix~\ref{app:ZNE_levels}.

When scaling the noise in our simulations we use factors $c=2,3$. To implement  $c=2$ we replace $XX(\theta)$ by $XX(\alpha)XX(\theta - \alpha)$ with $\alpha$ chosen randomly. We treat $R_Z$ and $R_Y$  in the same way.  To implement $c=3$ we replace $XX(\theta)$ by $XX(\theta)XX(\alpha)XX(-\alpha)$ performing an identity insertion. Again, we treat  $R_Z$ and $R_Y$  in the same way. Both procedures scale the number of all types of gates in our random quantum circuits by the noise level. 

\subsection{variable noise Clifford Data Regression  }

When constructing the noise-scaled circuits necessary for vnCDR we use the same method of noise scaling as used in the ZNE implementation.
In our simulations above we use  $c \in \{1,2,3\}$. For  the case of random quantum circuits, we  found that this choice of noise levels   results in  better results than using $c \in \{1,2\}$. The method used to construct the near-Clifford training circuits is discussed in Section~\ref{sec:training_set_construction}.

\subsection{Virtual Distillation}

We implement VD by computing  $\frac{{\rm Tr } [\rho^M X]}{{\rm Tr} [\rho^M]}$ for the noisy state $\rho$ and neglecting errors that occur  during the derangement circuit execution. This assumption enables classical simulation of VD for a $Q$-qubit system while simulating $Q$ noisy qubits instead of the $2Q+1$ that would otherwise be required to simulate the derangement. We analyze the performance  of VD with derangement errors for  $Q=4$ systems in Appendix \ref{app:swap_noise}. We present results of VD using $2$ copies as we observed that this typically gave the best results. We explore the performance of VD for various numbers of copies in Appendix~\ref{app:VD_copy_analysis}.

\subsection{Unified Data-driven Error Mitigation }
To implement UNITED we choose  the same near-Clifford circuits as used in vnCDR. The training set construction methods used to construct these circuits are discussed in Section~\ref{sec:training_set_construction}. We choose $M_{\rm max}=3$ and state preparation noise levels $c \in\{1,2,3\}$. We found that these parameters performed best across the regimes investigated for the RQC experiments. 

\subsection{Constructing the near-Clifford training circuits}
\label{sec:training_set_construction}
To construct the near-Clifford training circuits for the random quantum circuits we use an algorithm from~\cite{czarnik2021qubit}. We replace $XX$ and $R_Y$ gates with their closest Clifford gates of the same type. We also replace a fraction of $R_Z$ gates by their closest $R_Z$ Clifford gates, leaving $N=10$ $R_Z$ gates intact.

In order to maximize the information contained within the training set while minimizing the number of shots used to evaluate their outputs on noisy hardware we inspect the distribution of the exact values produced by these circuits. Initially, we construct $100$ near-Clifford circuits using the method described above. We find that with an increasing number of qubits, an increased fraction of their exact expectation values approaches $0$. This clustering effect is a property of near-Clifford quantum circuits~\cite{czarnik2022improving}. Shot efficiency of the learning-based error mitigation method can be improved by constructing a training set with significant variance in the exact expectation values, as explored in detail in~\cite{czarnik2022improving}. Therefore, to construct the training set in this work, we post-select $N_{t}=50$ near-Clifford circuits with the largest absolute value of the exact observable from the initial $100$. We find that this strategy does not decrease the quality of the mitigation while decreasing the shot cost, see Appendix~\ref{app:clustering} for details. 

However, the strategy outlined above is sub-optimal when applied to the Max-Cut QAOA problems, as for $Q=10$ we see instances for which all $100$ training circuits generated using this approach have the exact expectation value $0$. Therefore, we use a more advanced training set construction strategy. This strategy is based on the method proposed in~\cite{czarnik2022improving}. We use $N_t=100$ training circuits with $X^{\rm exact}$ evenly distributed between $-0.495$ and $0.495$, i.e. $X^{\rm exact} \in \{ -0.495 + 0.01n\}_{n=0}^{99}$. To construct these training circuits we start by building $10^5$ near-Clifford circuits using the same gate replacement strategy as outlined above, including the same number of non-Clifford gates  ($N=10$). We then post-select these circuits in order to match the desired distribution of exact expectation values. We wish to reinforce the point that this process is scalable as near-Clifford circuits are classically simulable. While previous studies used Markov-Chain Monte Carlo sampling to generate such distribution \cite{czarnik2021qubit}, the simpler method of post-selection can also be used to find the desired distribution in certain cases. 

\subsection{Shot budgeting}
\label{sec:budgeting}
Each technique described above requires a certain total number of circuits in order to mitigate the observable of interest. This, therefore, corresponds to a different number of observable evaluations. For VD, it is necessary to perform a measurement for  $2$ different circuits obtained with controlled derangement~\cite{koczor2020exponential}. In the case of ZNE, one needs to measure the expectation value of $n+1$ circuits built from one circuit of interest at $n+1$ noise levels. For vnCDR, we additionally need to evaluate each training circuit for $n+1$ noise levels. This means that in total vnCDR requires $(n+1)(N_{t}+1)$ circuit evaluations. Finally, for UNITED we need to mitigate the observable estimates using VD with $M_{\rm max}-1$ different copies for the training circuits, the circuit of interest, and for each of $n+1$  noise levels.  Therefore, the number of circuit evaluations required by UNITED is $(n+1)(N_{t}+1)(2M_{\rm max}-1)$.

In this work we consider several total shot budgets $N_{\rm tot}$. For each method, we distribute the shots across measurements  dividing $N_{\rm tot}$  by the number of measurements. We note that this strategy may not be optimal as it may be beneficial to distribute the shots unequally between noise levels or copy numbers. We note that one could imagine fine-tuning the values of  $n$, $N_{t}$, $M_{\rm max}$ for a given $N_{\rm tot}$ but leave the systematic investigation of more advanced budgeting strategies for future work. 

\section{Discussion}

The ability to efficiently and accurately mitigate hardware errors will be essential to achieving practical application of near-term quantum computers. A number of approaches have been proposed to fill this need, each bringing useful new ideas. In this work, we have discussed some state-of-the-art methods and examined how the approaches might best be combined. Specifically, we have considered Zero Noise Extrapolation (ZNE), variable noise Clifford Data Regression (vnCDR), and Virtual Distillation (VD) and proposed two new methods that represent unifications of the ideas behind these approaches. We applied these methods to the mitigation of errors in local observables measured from states prepared with random circuits and observables produced from applying QAOA to Max-Cut problems. As these methods have different advantages, disadvantages, and resource requirements, we evaluated their performance with $10^5-10^{10}$ total shots to mitigate the expectation value of an observable of interest with up to $Q=10$ qubits with depth $L=Q$ as well as up to $Q=8$ for depth $L=16Q$. We used a realistic simulation of a noisy trapped-ion device, which was chosen as the trapped-ion architecture enables all-to-all connectivity.

Combining the near-Clifford circuit data and the multi-copy data from VD, we formulated Clifford-guided VD (CGVD) which uses Clifford data to guide a combination of VD-mitigated observables to obtain an improved estimate. We then generalized this method to perform a Clifford-guided Richardson-like extrapolation on VD-mitigated data at variable noise levels. This can be thought of as first suppressing the effect of orthogonal errors, by using a combination of VD-mitigated observables, and then extrapolating to the zero noise limit. The new method, UNIfied Technique for Error mitigation with Data (UNITED), can be viewed as a unification of the CDR, ZNE, and VD methods. Furthermore, vnCDR is a special case of the new general method.  

We find that UNITED is the technique that offers the best mitigation in both the RQC and QAOA numerical experiments at $N_{\rm tot}=10^{10}$. We also find that vnCDR is preferable at budgets $N_{\rm tot}=10^6,10^8$ in the case of RQC. For $N_{\rm tot}=10^{5}$, ZNE is optimal for circuits  with depths $L=Q$ while vnCDR is optimal for more noisy $L=16Q$. When considering the QAOA results, VD is optimal for all shot budgets apart from the largest for which UNITED is the best method. This is because the noise floor is significantly lower in QAOA due to the fact that the solution of the algorithm is also an eigenstate of the measurement operator \cite{koczor2021dominant}. We observe that increasing the shot budget improves the error-mitigated expectation value until a performance ceiling is reached. Different techniques have different performance ceilings and we find that UNITED, being able to draw from more data than VD and ZNE, reaches better mitigated values when more resources are available. 

We note it may be difficult to assess when the performance ceiling of a given method has been reached. This is an issue particularly for methods that lack analytical scaling guarantees, or when such guarantees only exist under restrictive noise assumptions. In this case, we propose using the convergence of the mitigated observable with the number of shots employed to judge if the performance ceiling has been reached.

As the available quantum devices continue to grow in qubit number and reduce the rate of hardware errors, practical applications of quantum computing grow ever closer. Due to the high number of qubits that are required for full error correction, it seems likely that error mitigation will play a pivotal part in those early applications. This work provides a useful reference for those who are seeking such applications to determine which combination of strategies is best for their use case. 

A natural question one might ask is if there are other forms of data that would be useful to include in an error mitigation approach. Error mitigation for deep quantum circuits is challenging, but perhaps some other source of data (other than those we considered) would extend the depth for which error mitigation is practical. Further unification with other error mitigation techniques such as noise-resilient quantum compiling \cite{cincio2018learning,cincio2021machine,murali2019noise,khatri2019quantum,sharma2019noise}, quasi-probabilistic error decomposition~\cite{temme2017error,takagi2020optimal},  verified phase estimation \cite{o2020error}, truncated Neumann series~\cite{wang2021mitigating} and specific approaches leveraging symmetries and/or post-selection ~\cite{mcardle2019error,bonet2018low,otten2019noise,cai2021quantum} is likely to be a promising research direction. 

Another interesting direction for future work is to consider how best to determine when to measure new expectation values (i.e. more near-Clifford circuits, more noise levels, more copies of the state, etc.) versus expanding more shots per expectation value in order to get the best mitigation possible with fixed resources. Finally, we note that investigation of the effects of noisy controlled derangement  on VD and UNITED performance including large circuits challenging for noisy classical simulation  will be necessary for a full understanding of these methods' potential. 

\section{Acknowledgments}

This work was supported by the Quantum Science Center (QSC), a National Quantum Information Science Research Center of the U.S. Department of Energy (DOE). DB was supported by the U.S. DOE through a quantum computing program sponsored by the Los Alamos National Laboratory (LANL) Information Science \& Technology Institute and by the European Union’s Horizon 2020 research and innovation program under the Marie Sk\l{}odowska-Curie grant agreement No. 955479. M.H.G was supported by “la Caixa” Foundation (ID100010434), Grant No. LCF/BQ/DI19/11730056. Piotr C. and AA were supported by the Laboratory Directed Research and Development (LDRD) program of LANL under project numbers 20190659PRD4 (Piotr C.) and 20210116DR (AA and Piotr. C). MC  was supported by the U.S. Department of Energy, Office of Science, Office of Advanced Scientific Computing Research, under Computational Partnerships program. PJC also acknowledges initial support from the LANL ASC Beyond Moore's Law project. LC was also initially supported by the U.S. DOE, Office of Science, Office of Advanced Scientific Computing Research, under the Quantum Computing Application Teams~program.

\bibliography{main_quantum}

\appendix

\section{Cancellation of errors using Richardson extrapolation}

\label{app:Richardson}

By using Richardson extrapolation one can cancel the leading orders of a Taylor expansion of a noisy observable expectation value  $\hat{\mu}^{\rm ZNE}(\lambda)$  to suppress its error. The order to which one can suppress the error depends on how many noisy observables one has access to. Each observable used must be evaluated with a different noise strength, which should be known. Here $\lambda$ is the noise strength. More precisely, Richardson extrapolation uses a combination of $n+1$ noisy expectation values of the mitigated observable $\hat{\mu}^{\rm ZNE}_{0}, \hat{\mu}^{\rm ZNE}_{1},\dots,\hat{\mu}^{\rm ZNE}_{n}$ with the noise strengths 
$\lambda_0<\lambda_1<\dots<\lambda_n$  to suppress the error up to $\mathcal{O}(\lambda_0^{n+1})$. 
Writing the series expansion of the noisy observable as a function of $\lambda$ we have 
\begin{equation}
\hat{\mu}(\lambda) = \mu + \sum_{i=1}^{\infty} \alpha_i \lambda^i,
\end{equation}
where $\mu$ is the exact expectation value and $\alpha_i$ are the expansion coefficients. We introduce the noise levels $c_0, c_1,c_2,\dots,c_n$ such that  $\lambda_0,\lambda_1, \lambda_2, ..., \lambda_n = c_0\lambda_0,c_1\lambda_0, c_2\lambda_0, ..., c_n \lambda_0$, which enables us to write
\begin{equation}
\hat \mu_j^{\rm ZNE} = \hat{\mu}(c_j\lambda_0) = \mu + \sum_{i=1}^{\infty} \alpha_i c_j^i\lambda_0^i,
\end{equation}

To build intuition, we first consider the combination of just two noisy observables, $\hat{\mu}^{\rm ZNE}_0$ and $\hat{\mu}^{\rm ZNE}_1$, evaluated at noise levels $c_0$ and $c_1$, respectively. We show how by using Richardson extrapolation we can cancel the error in an estimate of $\mu$ up to order $\mathcal{O}(\lambda_0^{2})$. To begin with, we write the series expansion for $\hat{\mu}^{\rm ZNE}_0$ and $\hat{\mu}^{\rm ZNE}_1$ as
\begin{align}
    \hat{\mu}^{\rm ZNE}_0 &= \mu + \alpha_0\lambda_0 + \mathcal{O}(\lambda_0^2), \\
    \hat{\mu}^{\rm ZNE}_1 &= \mu + \alpha_0 c_1 \lambda_0 + \mathcal{O}(\lambda_0^2).
\end{align}
Note that we use $c_0=1$ above.
Therefore, combining $\hat{\mu}^{\rm ZNE}_0$ and $\hat{\mu}^{\rm ZNE}_1$ with coefficients $\gamma_0$ and $\gamma_1$ we write
\begin{align}
\label{richardson_cancelling}
   \mu^{\rm ZNE}_1 &=  \gamma_0 \hat{\mu}_0 + \gamma_1 \hat{\mu}_1 \nonumber \\
   &= (\gamma_0 + \gamma_1)\mu + \alpha_0\lambda_0(\gamma_0 + c_1 \gamma_1) + \mathcal{O}(\lambda_0^2),
\end{align}
Therefore, to cancel terms up to order $\mathcal{O}(\lambda_0^2)$ we require:
\begin{align}
   & \gamma_0 + \gamma_1 = 1, \\
   & \gamma_0 + c_1\gamma_1 = 0.
\end{align}

One  can  generalize this procedure to larger numbers of noise levels. In such a case we  take a combination of $n+1$ $\hat{\mu}^{\rm ZNE}_j$ observables
\begin{equation}
\mu^{\rm ZNE}_n = \sum_{j=0}^{n} \gamma_j \hat{\mu}^{\rm ZNE}_{j},
\end{equation}
with $\gamma_j$ being the coefficients. 
Expanding this sum in powers of $\lambda_j$ gives
\begin{align}
\mu^{\rm ZNE}_n &= \sum_{j=0}^{n} \gamma_j \mu + \sum_{j=0}^{n} \gamma_j  \sum_{i=1}^{\infty} \alpha_i \lambda_j^i \nonumber\\
&= \sum_{j=0}^{n} \gamma_j \mu +  \sum_{i=1}^{\infty} \alpha_i \sum_{j=0}^{n} \gamma_j \lambda_j^i.   
\end{align}
Using the noise levels  we can write above as 
\begin{equation}
\mu^{\rm ZNE}_n  = \sum_{j=0}^{n} \gamma_j \mu +  \sum_{i=1}^{\infty} \alpha_i  \lambda_0^i \sum_{j=0}^{n} \gamma_j c_j^i.
\label{eq:exp_Rich}
\end{equation}
Therefore, to obtain the error of order  $\mathcal{O}(\lambda_0^{n+1})$ the Richardson extrapolation chooses $\gamma_j$ coefficients to cancel the leading $n$ orders of the error  in \eqref{eq:exp_Rich}. This leads to a system of linear equations for $\gamma_j$
\begin{equation}
\sum_{j=0}^{n} \gamma_j =1, \quad \sum_{j=0}^{n} \gamma_j c_j^i = 0 \, \ {\rm for} \, i=1,\ldots,n.
\end{equation}
To sum up, the Richardson extrapolation uses a combination of different noisy estimates with different noise levels  to mitigate the error by canceling the leading error terms.

\section{Suppressing error through a combination of Virtual-Distillation-mitigated observables}
\label{app:VD_deriv}

Here we consider a combination of VD-mitigated observables  $\hat{\mu}^{\rm VD}_M$  obtained for different numbers of VD copies  $M=1,2,\dots ,M_{\rm max}$. 
By analogy to the Richardson extrapolation, we look for a combination
\begin{equation}
\mu'_{M_{\rm max}} = \sum_{M=1}^{M_{\rm max}}b_{M}\hat{\mu}^{\rm VD}_{M}
\label{eq:M_comb}
\end{equation}
with the  error smaller than the error of each of  the individual observables. As proved below, Virtual Distillation suppresses contributions to the observable from eigenstates of the noisy state other than the dominant one leading to 
\begin{equation}
|\hat{\mu}^{\rm VD}_M - \mu  - \epsilon | \in   \mathcal{O}(\xi^M), 
\label{eq:VD_suppr}
\end{equation}
where $\mu$ is the exact observable, $\epsilon$ is the noise floor defined in \eqref{eq:noise_floor} and repeated below for convenience
\begin{equation}
\epsilon = \langle \psi_0 | X | \psi_0 \rangle - \langle \psi_{\rm exact} | X | \psi_{\rm exact} \rangle,
\end{equation} and 
\begin{equation}
\xi = \sum_{i=1}^{2^Q-1} p_i = 1 -p_0
\label{eq:xi}
\end{equation}
is a sum of  eigenvalues of the mitigated state \eqref{eq:eigen} other than the dominant one.

\subsection{Proof of  VD error scaling with the number of state copies}

To prove (\ref{eq:VD_suppr}) we use the eigendecomposition of a noisy state $\rho$ given by Eq.~\eqref{eq:eigen}. Then we have
\begin{equation}
\hat{\mu}^{\rm VD}_{M} = \frac{{\rm Tr } [\rho^M X]}{{\rm Tr} [\rho^M]} = \frac{\sum_{i=0}^{2^Q-1}p_i^M X_i}{\sum_{i=0}^{2^Q-1}p_i^M},
\end{equation}
where  $X_i=\langle\psi_i|X|\psi_i\rangle$. Transforming it further, we obtain
\begin{align}
\hat{\mu}^{\rm VD}_{M} &=  \frac{p_0^M X_0+\sum_{i=1}^{2^Q-1}p_i^M X_i}{p_0^M+\sum_{i=1}^{2^Q-1}p_i^M} \nonumber \\
&= X_0 + \frac{\sum_{i=1}^{2^Q-1}p_i^M (X_i-X_0)}{p_0^M+\sum_{i=1}^{2^Q-1}p_i^M}.
\end{align}
Therefore, we obtain
\begin{align}
|\hat{\mu}^{\rm VD}_{M} - X_0| =  \frac{\sum_{i=1}^{2^Q-1}p_i^M |X_i-X_0|}
{p_0^M+\sum_{i=1}^{2^Q-1}p_i^M} \le  \frac{2\lVert X\rVert_{\infty}\sum_{i=1}^{2^Q-1}p_i^M }
{p_0^M+\sum_{i=1}^{2^Q-1}p_i^M}, 
\label{eq:lim1}
\end{align}
where we use that $p_i \ge0$, and $\lVert X\rVert_{\infty}$ is an infinity norm of $X$.  But, 
\begin{equation}
\frac{\sum_{i=1}^{2^Q-1}p_i^M }
{p_0^M+\sum_{i=1}^{2^Q-1}p_i^M} \le \frac{\sum_{i=1}^{2^Q-1}p_i^M }{p_0^M} \le \frac{(\sum_{i=1}^{2^Q-1}p_i)^M }{p_0^M},
\label{eq:lim2}
\end{equation}
where we again use that $p_i \ge0$. Using (\ref{eq:lim1}),  (\ref{eq:lim2}), and (\ref{eq:xi}), 
this leads to
\begin{equation}
|\hat{\mu}^{\rm VD}_{M} - X_0| \le \frac{2\lVert X\rVert_{\infty} \xi^M}{(1-\xi)^M}.
\end{equation}
Therefore, if $\lVert X\rVert_{\infty}$ is finite, like for a Pauli string, we obtain 
\begin{equation}
|\hat{\mu}^{\rm VD}_{M} - X_0|  = |\hat{\mu}^{\rm VD}_{M} - \mu - \epsilon| \in \mathcal{O}(\xi^M),
\label{eq:VD_suppr2}
\end{equation}
that was to be proved.

\subsection{Error suppression with a combination of VD terms}
\label{app:VD_comb}

Here, to motivate the CGVD ansatz \eqref{eq:CGVD_ansatz} - which is a linear combination of VD-mitigated observables with up to  $M_{\rm max}$ copies - we  find a combination of VD-mitigated observables that estimates the noise-free observable of interest with an error smaller than its parts. We start with $M_{\rm max}=2$ and then generalize this approach to more copies. 

The noise model we consider is a simple model where for each gate an error occurs with probability $\lambda$, and every two different sequences of errors result in orthogonal states.
Additionally, at the end of the circuit we place a coherent error given by a unitary $U(\theta) = e^{-i \theta E}$, where $E$ is a Hermitian operator. Such an error produces a non-zero noise floor for $\theta \neq 0$, while preserving the orthogonality of states corresponding to different sequences of the gate errors.  For a circuit with $G$ gates the noisy output $\rho$ can be written as
\begin{align}
    \rho &= (1-\lambda)^{G} \rho_{0} + (1-\lambda)^{G-1} \lambda \sum_{j=1}^{G} \rho_{j} \nonumber \\
         &+ (1-\lambda)^{G-2}\lambda^2\sum_{j_1 \neq j_2} \rho_{j_1, j_2} + \cdots \ ,
\end{align}
where every state is orthogonal. For small enough error rate, $\rho_{0}$ is the dominant eigenvector, i.e. $\rho_{0} = |\psi_{0} \rangle \langle \psi_{0} |$. 
We consider taking linear combinations of observables mitigated with VD using different numbers of copies up to $M=M_{\rm max}$. That is,
\begin{equation}
\mu'_{M_{\rm max}} = \sum\limits_{M=1}^{M_{\rm max}}b_{M}\hat{\mu}^{\rm VD}_{M} \ ,
\label{eq:comb_VD}
\end{equation}
where $b_{M}$ are some coefficients to be found. We show that this ansatz contains solutions that lead to an estimate of the observable of interest with a smaller error than $\hat{\mu}^{\rm VD}_{M_{\rm max}}$.

\subsubsection{Combining two VD-mitigated observables}
To build intuition we first consider the combination of VD-mitigated observables with $M=1$ and $M=2$ copies. We can write
\begin{align}
    \hat{\mu}^{\rm VD}_1 &= \frac{\tr[\rho X]}{\tr[\rho]} \nonumber\\
                &= (1-\lambda)^G X_{0} + (1-\lambda)^{G-1} \lambda \sum_{j=1}^{G} \tr[\rho_j X] + \cdots \nonumber\\
                &= \big[1-G\lambda + 1/2(G-1)G\lambda^{2}+\cdots\big]X_0 \nonumber\\
                &\quad + \big[1 - (G-1)\lambda + \nonumber\\
                &\quad + 1/2(G-2)(G-1)\lambda^{2} + \cdots \big]\lambda \sum_{j=1}^G \tr[\rho_{j} X] \nonumber\\
                &\quad + \cdots.
\end{align}
By grouping terms according to powers of $\lambda$ we rewrite this as 
\begin{equation}
    \hat{\mu}^{\rm VD}_1 = X_{0} + \lambda \alpha_1^{(1)} + \lambda^{2} \alpha_{2}^{(1)} + \lambda^{3} \alpha_{3}^{(1)} + \cdots,
\end{equation}
where we have defined $ \alpha_1^{(1)}=\sum_{j=1}^{G}\tr[\rho_{j}X]-GX_0$, and $\alpha_{2}^{(1)}=1/2(G-1)GX_0-(G-1)\sum_{j=1}^{G}\tr[\rho_{j}X]+\sum_{j_1 \neq j_2}\tr[\rho_{j_1,j_2}X]$. In the following we assume that $\alpha_1^{(1)}\neq 0$ so that error is of order $\mathcal{O}(\lambda)$.
Taking into account that 
\begin{equation}
\tr [\rho^2] =\big[(1-\lambda)^2 + \lambda^2\big]^G
\end{equation}
we arrive at a similar expression for $\hat{\mu}^{\rm VD}_2$, 
\begin{align}
    \hat{\mu}^{\rm VD}_2 &= \frac{(1-\lambda)^{2G} X_{0} + (1-\lambda)^{2(G-1)}\lambda^{2} \sum_{j=1}^{G} \tr[\rho_j X] + \cdots}{\big[(1-\lambda)^2 + \lambda^{2}\big]^{G}} \nonumber\\
    & =  X_{0} + \lambda^{2} \alpha_1^{(2)} + \lambda^{3} \alpha_{2}^{(2)} + \cdots,
\end{align}
where the coefficients $\alpha_k^{(2)}$ are defined similarly to the $\alpha_k^{(1)}$  ones. In particular, one can find $\alpha_1^{(2)}=\alpha_1^{(1)}$.  As expected, when computing the expectation value with two copies of the state, Virtual Distillation suppresses the error contribution from the orthogonal noise of order $\mathcal{O}(\lambda)$. 

Let us now show how one can take a linear combination of  $\hat{\mu}^{\rm VD}_1$ and $\hat{\mu}^{\rm VD}_2$ to further suppress the error. Consider the combination 
\begin{equation}
    \mu_{2}' = b_1 \hat{\mu}^{\rm VD}_1 + b_2 \hat{\mu}^{\rm VD}_2,
\end{equation}
where $b_1$ and $b_2$ are coefficients to be found. Expanding this out we obtain
\begin{align} \label{eq:mu-prime}
    \mu_2' &= X_{0} \big( b_1  + b_2\big) + \big(\lambda \alpha_{1}^{(1)}b_1 + \lambda^{2} \alpha_{1}^{(2)}b_{2}\big) \\
                 &\quad + \lambda^{2} \alpha_{2}^{(1)} b_1 + \cdots.\nonumber
\end{align}
Consider taking
\begin{align}
    &b_1 + b_2 =1 \\
    & \lambda \alpha_{1}^{(1)}b_1 + \lambda^{2} \alpha_{1}^{(2)}b_{2} = 0.
\end{align}
Solving for $b_1$ and $b_2$, and recalling that $\alpha_{1}^{(2)}=\alpha_{1}^{(1)}$  gives
\begin{align}
    b_1 &= \frac{-\lambda \alpha_{1}^{(2)}}{\alpha_{1}^{(1)} - \lambda\alpha_{1}^{(2)}}=\frac{-\lambda }{(1- \lambda)}\,, \\
    b_2 &= \frac{\alpha_{1}^{(1)}}{\alpha_{1}^{(1)} - \lambda\alpha_{1}^{(2)}}= \frac{1}{(1- \lambda)}\,.
\end{align}
Putting these coefficients  into Eq.~\eqref{eq:mu-prime}  leads to 
\begin{equation}
    \mu_{2}' = X_{0} + \lambda^{2} \alpha_{2}^{(1)}b_1 + \cdots.
\end{equation}
Considering the term $ \lambda^{2} \alpha_{2}^{(1)}b_1$ we obtain
\begin{equation}
      \lambda^{2} \alpha_{2}^{(1)}b_1   = -\lambda^{3} \frac{  \alpha_{2}^{(1)}}{1 - \lambda}\in  \mathcal{O}(\lambda^{3})\,.
\end{equation}
The previous then implies 
\begin{equation}
  | \mu_{2}' - X_0| = | \mu_{2}' - \mu - \epsilon| \in \mathcal{O}(\lambda^{3}).
\end{equation}
Therefore, we have shown that taking a linear combination of VD-mitigated observables further reduces the error on the estimate of the noiseless observable. Having seen how the errors can be canceled for $M_{\rm max} =2$ let's generalize this same error cancellation to the many copy case. 

\subsubsection{Generalizing to the many copy case}

Analogously to the case of $M_{\rm max}=2$ one can use  combination~\eqref{eq:comb_VD} to suppress the error also for larger $M_{\rm max}$. For $M$ copies and our simple noise model we have 
\begin{align}
    & \hat{\mu}^{\rm VD}_M = \frac{\tr[\rho^M X]}{\tr[\rho^M]} = \nonumber \\ &  \frac{(1-\lambda)^{MG} X_{0} + (1-\lambda)^{M(G-1)}\lambda^{M} \sum_{j=1}^{G} \tr[\rho_j X] + \cdots}{\big[(1-\lambda)^M + \lambda^{M}\big]^{G}} \nonumber\\
    & =  X_{0} + \lambda^{M} \alpha_1^{(M)} + \lambda^{M+1} \alpha_{2}^{(M)} + \cdots,
\end{align}
where $\alpha_1^{(M)}$, $\alpha_2^{(M)}$ are coefficients of  $\hat{\mu}^{\rm VD}_M$  series expansion in $\lambda$. 
Using that and grouping  terms based on the index~$j$ in $\alpha_{j}^{(M)}$ we obtain 
\begin{align}
    \mu_{M_{\rm max}}' &= \sum\limits_{M=1}^{M_{\rm max}}b_{M}\hat{\mu}^{\rm VD}_{M} \nonumber \\
    &=  X_{0}  \sum_{M=1}^{M_{\rm max}}  b_M + \sum_{M=1}^{M_{\rm max}} b_M \sum_{j=1}^{\infty} \alpha_j^{(M)}  \lambda^{j+M-1} = \nonumber  \\
     &=  X_{0}  \sum_{M=1}^{M_{\rm max}}  b_M +
     \sum_{j=1}^{\infty} \sum_{M=1}^{M_{\rm max}} b_M 
     \alpha_j^{(M)}  \lambda^{j+M-1} = \nonumber \\
      &=  X_{0}  \sum_{M=1}^{M_{\rm max}}  b_M +
     \sum_{j=1}^{M_{\rm max}-1} \sum_{M=1}^{M_{\rm max}} b_M 
     \alpha_j^{(M)}  \lambda^{j+M-1}  +  \nonumber \\
     &+b_1 \alpha_{M_{\rm max}}^{(1)} \lambda^{M_{\rm max}}   + \mathcal{O}(\lambda^{M_{\rm max+1}}).
\end{align}
Taking 
\begin{align}
     \sum_{M=1}^{M_{\rm max}}  b_M &= 1  \label{eq:gammaM1}, \\
     \sum_{M=1}^{M_{\rm max}} b_{M} \alpha_j^{(M)}  \lambda^{j+M-1} &=0\,, \quad \forall j=1,...,M_{\rm max} -1\,,  \label{eq:gammaM2}
\end{align} 
we get 
\begin{equation}
    \mu_{M_{\rm max}}' = X_0 +   b_1 \alpha_{M_{\rm max}}^{(1)} \lambda^{M_{\rm max}} + \mathcal{O}(\lambda^{M_{\rm max+1}}) \ .
\end{equation}

To obtain  (\ref{eq:gammaM2}) in the limit of small $\lambda$ while taking into account  that the $\alpha_j^{(M)}$ coefficients do not depend on $\lambda$ we need terms $b_1$, $b_2 \lambda$, $b_3 \lambda^2$,\dots,$b_{M_{\rm max}} \lambda^{M_{\rm max}-1}$ to be of the same order.
Otherwise, a system of equations (\ref{eq:gammaM1}, \ref{eq:gammaM2}) could be reduced to systems of overdetermined linear equations which usually do not have a solution.
Furthermore,  Eqs.~\ref{eq:gammaM1},~\ref{eq:gammaM2} imply that $b_{M_{\rm max}}$ is of order~$1$, which leads to
$b_{1} \in \mathcal{O}(\lambda^{M_{\rm max}-1})$. Assuming that the system of equations (\ref{eq:gammaM1}, \ref{eq:gammaM2}) has a solution, which is only false in the special case of a singular system of linear equations, we obtain
\begin{align}
    | \mu_{M_{\rm max}}' - X_0| = | \mu_{M_{\rm max}}' - \mu - \epsilon| \in  \mathcal{O}(\lambda^{M_{\rm max} + 1}).
\end{align}
Therefore, the combination improves the VD error suppression for the largest used $M=M_{\rm max}$. 

This shows that in general, we can combine many VD-mitigated observables with different numbers of copies to  achieve a better quality of error mitigation. Consequently, it  motivates the linear combinations of VD-mitigated observables we take in the CGVD~\eqref{eq:CGVD_ansatz} and UNITED~\eqref{eq:UNITED_ansatz} Ans{\"a}etze proving that such  combinations can be advantageous. 
We note that the coefficients $b_M$ given by the system of equations (\ref{eq:gammaM1}, \ref{eq:gammaM2}) are not guaranteed to be a unique  choice that leads to an improvement of the error of combination (\ref{eq:comb_VD}) with respect to its constituent parts.  

\subsection{Combining  VD-mitigated observables obtained at different noise levels}
\label{app:VD_noise_comb}
Now we motivate the ansatz used in the UNITED method, which takes the form of a linear combination of VD-mitigated observables obtained with different numbers of copies and different noise levels.  We consider the same noise model as introduced in the previous section (Appendix~\ref{app:VD_comb}), where we showed that the orthogonal error can be suppressed by taking a combination of the VD-mitigated observables with different numbers of copies. Such a combination in itself cannot mitigate a noise floor that results from a coherent error of the form $U(\theta) = e^{-i \theta E}$. To mitigate the noise floor we consider taking a combination of VD-mitigated observables obtained  at various noise levels. 

We start by recalling from the previous section that
\begin{equation}
    \hat{\mu}^{\rm VD}_{M} = X_{0} + \lambda^{M} \alpha_{1}^{(M)} + \lambda^{M+1} \alpha_{2}^{(M)} + \cdots
\end{equation}
and 
\begin{equation}
    X_0 = \langle \psi_0 | X | \psi_0 \rangle \ ,
\end{equation}
where $|\psi_0\rangle\langle \psi_0|$ is the dominant eigenvector of the noisy state $\rho$. For our noise model we have
\begin{equation}
    X_0 = \tr[e^{-i \theta E}|\psi_{\rm exact}\rangle\langle\psi_{\rm exact}|e^{i \theta E}X] \ ,
\end{equation} 
where we assume $[X, E] \neq 0$. This leads to
\begin{align}
    \hat{\mu}^{\rm VD}_{M}(\theta, \lambda) &= \tr[e^{-i  \theta E}|\psi_{\rm exact}\rangle\langle\psi_{\rm exact}|e^{i  \theta E}X] +  \lambda^{M} \alpha_{1}^{(M)} \nonumber \\ 
    & \quad + \lambda^{M+1} \alpha_{2}^{(M)} + \cdots.
\end{align}
As shown in the previous section, combining VD-mitigated observables as 
\begin{equation}
\mu'_{M_{\rm max}}(\theta, \lambda) = \sum\limits_{M=1}^{M_{\rm max}}b_{M}\hat{\mu}^{\rm VD}_{M}(\theta, \lambda) \ ,
\label{eq:comb_VDp}
\end{equation}
with coefficients (\ref{eq:gammaM1}, \ref{eq:gammaM2})
leads to 
\begin{align}
        \mu'_{M_{\rm max}}(\theta, \lambda) &= \tr[e^{-i  \theta E}|\psi_{\rm exact}\rangle\langle\psi_{\rm exact}|e^{i  \theta E}X] \nonumber \\
        & \quad + \mathcal{O}(\lambda^{M_{\rm max} +1}).
\end{align}

For large enough $M_{\rm max}$ the dominant part of the error comes from the noise floor. Therefore, in this case we can neglect  dependence  of $\mu'_{M_{\rm max}}$ on $\lambda$ 
\begin{equation}
     \mu'_{M_{\rm max}}(\theta, \lambda) \approx 
     \tr[e^{-i  \theta E}|\psi_{\rm exact}\rangle\langle\psi_{\rm exact}|e^{i  \theta E}X]. 
\end{equation}
Performing a Taylor expansion of this approximation around $\theta=0$, we obtain 
\begin{equation}
    \mu'_{M_{\rm max}}(\theta)  = \mu + \beta_{1}  \theta + \beta_2  \theta^{2} + \cdots,
    \label{eq:comb_VDpapprox}
\end{equation}
where $\mu = \tr[|\psi_{\rm exact}\rangle\langle\psi_{\rm exact}|X]$, and $\beta_1,\beta_2,\dots$ are the expansion coefficients. By considering $n+1$ values of $\theta$, i.e.  $c_0\theta_0 =\theta_0$, $c_1 \theta_0$, $\dots$, $c_n\theta_0$,   we perform a Richardson extrapolation obtaining a mitigated estimate of the observable 
\begin{align}
    \mu''_{M_{\rm max},n} &= \sum_{j=0}^{n} \gamma_j \mu'_{M_{\rm max}}(c_j\theta_0) \ ,
\label{Richardson_Mismatch} 
\end{align}
where 
\begin{equation}
\sum_{j=0}^{n} \gamma_j =1, \quad \sum_{j=0}^{n} \gamma_j c_j^k = 0 \, \ {\rm for} \, k=1,\ldots,n.
\end{equation}

Recalling (\ref{eq:comb_VDp}), (\ref{eq:comb_VDpapprox}) we finally obtain  
\begin{equation}
    \mu''_{M_{\rm max}, n} =  \sum_{j=0}^{n}  \sum\limits_{M=1}^{M_{\rm max}} \gamma_j b_{M} \hat{\mu}^{\rm VD}_{M}(c_j\theta_0, \lambda) \ ,
\end{equation}
which is a linear combination of VD-mitigated observables obtained with different numbers of copies at different coherent  noise levels.  This combination has exactly the same form as the UNITED ansatz~(\ref{eq:UNITED_ansatz}). As previously discussed in Appendix~\ref{app:Richardson} this application of the Richardson extrapolation results in an error
\begin{equation}
    \mu''_{M_{\rm max}, n} - \mu = \mathcal{O}(\theta_0^{n+1}),
    \label{eq:err_scal}
\end{equation}
assuming that the orthogonal errors are small enough. We note that as long as the orthogonal error rates are not too large, one can scale both $\theta$ an $\lambda$, e.g. choosing  $\theta_j = c_j \theta_0$ and $\lambda_j = c_j \lambda_0$, which results in
\begin{equation}
    \mu''_{M_{\rm max}, n} =  \sum_{j=0}^{n}  \sum\limits_{M=1}^{M_{\rm max}} \gamma_j b_M \hat{\mu}^{\rm VD}_{M}(c_j\theta_0, c_j \lambda_0),
\end{equation}
and obtaining the same error scaling as in (\ref{eq:err_scal}), since for large enough $M_{\rm max}$ the error will always be dominated by the coherent mismatch.

\section{Correcting for global depolarizing noise with CGVD or UNITED}
\label{app:CGVD_global}
Here we motivate the UNITED ansatz by considering its performance in mitigating the action of a global depolarizing noise channel on some observable of interest. The ansatzes used in CDR and vnCDR have been shown to perfectly correct global depolarizing noise and this fact is used to motivate the form of the CGVD ansatz. We note in this setting there is no limit set by coherent mismatch when using VD, so one expects a combination of VD and CDR to perform well. First, we consider CGVD since the treatment is the same as in the UNITED case and can be generalized very simply.

Consider a global depolarizing noise channel that acts on all $Q$ qubits. For some state $\rho = \ket{\psi}\bra{\psi}$ produced by quantum circuit $U$, this channel takes the form:
\begin{equation}
    \rho \longrightarrow (1-\lambda)\rho +\lambda \frac{\mathbb{1}}{2^Q}.
    \label{eq:dep_noise_channel}
\end{equation}
Consider the effect of the above channel on some observable of interest $X$
\begin{align}
    \hat{\mu} =(1-\lambda) \mu + \lambda\frac{\tr(X)}{2^Q},
\end{align}
where $\mu = \tr{(\rho X)}$. We consider the CGVD ansatz that combines observables evaluated at several numbers of copies in the VD circuit structure. For simplicity, we omit the small changes in the rate of errors for different numbers of copies due to the swap noise. We can rewrite~\eqref{eq:CGVD_ansatz} as follows
\begin{align}
    \mu'_{M_{\rm max}} = \sum_{M=1}^{M_{\rm max}} b_{M} \hat{\mu}^{\rm VD}_{M},
\label{eq:CGVD_ansatz_2}
\end{align}
where $\mu'_{M_{\rm max}}$ is the estimate of the observable of interest,  $a_{M}$ are the parameters fitted using least squares regression on the data generated by the near-Clifford training circuits mitigated using VD with up to $M$ copies. 

We can express an observable mitigated with VD using $M$ copies as 
\begin{equation}
    \hat{\mu}^{\rm VD}_{M} = \frac{\tr{\big[\rho^{M}X}\big]}{\tr{\big[\rho^{M}}\big]}.
\label{eq::VD}
\end{equation}
We can evaluate $\hat{\mu}^{\rm VD}_{M}$, assuming $\rho$ has been acted upon by a global depolarizing channel as defined in \eqref{eq:dep_noise_channel}. First focusing on the denominator,
\begin{align}
    \tr{\big[\rho^{M}\big]} &= \tr{\bigg[\sum_{k = 0}^{M}\begin{pmatrix}M\\k\end{pmatrix}\big((1-\lambda)\rho\big)^{(M-k)}\bigg(\frac{\lambda}{2^Q}\bigg)^{k}\bigg]} \nonumber \\
     &= \sum_{k = 0}^{M}\begin{pmatrix}M\\k\end{pmatrix}(1-\lambda)^{(M-k)}\tr{(\rho}^{M-k})\bigg(\frac{\lambda}{2^Q}\bigg)^{k} \ .
\end{align}
Noting that when $M=k$ we are tracing over the identity, we obtain:
\begin{align}
    \tr{\big[\rho^{M}\big]} = \bigg(\frac{1+(2^Q-1)(1-\lambda)}{2^Q}\bigg)^{M}+(2^Q-1)\bigg(\frac{\lambda}{2^Q}\bigg)^{M} \ .
\end{align}
Now evaluating the numerator of~\eqref{eq::VD} we obtain
\begin{align}
    \begin{split}
            \tr{\big[\rho^{M} X\big]}  =\tr{[\rho X]}\bigg(\frac{1+(2^Q-1)(1-\lambda)}{2^Q}\bigg)^{M} \\ 
            + \bigg(\frac{\lambda}{2^Q}\bigg)^{M} \bigg(\tr{[X]} - \tr{[\rho X]} \bigg) \ .
    \end{split}
\end{align}
Combining the above we arrive at the following expression,
\begin{align}
\hat{\mu}^{\rm VD}_{M} = f_{M}\tr{[\rho X]}+\frac{1-f_M}{2^Q}\tr{[X]}
\label{eq:dep_noise_eff} \ ,
\end{align}
where
\begin{align}
    f_{M} = 1 - \frac{2^Q \lambda^M}{(2^Q - 1) \lambda^M + ( 2^Q - \lambda(2^Q - 1 ))^M}
\end{align}
Taking $M=1$ leads to $f_{1} = 1-\lambda$ which gives the expected result. Using~\eqref{eq:CGVD_ansatz_2} we can determine the criteria for this ansatz to give perfect mitigation,
\begin{equation}
    \mu'_{M_{\rm max}} = \sum_{M=1}^{M_{\rm max}} b_{M} \big( f_{M} \tr{[\rho X]} + \frac{1 - f_M}{2^Q}\tr{[X]}\big).
\label{eq:perf_Mit_CGVD}
\end{equation}
Therefore, we require 
\begin{equation}
    \sum_{M=1}^{M_{\rm max}} b_{M} f_{M} = 1,  \quad \tr{[X]} = 0.
\end{equation}
A global depolarizing noise channel will have the same effect on the training circuits as it will on the circuit of interest. As we have access to the exact observable for these circuits, the coefficients required to lead to perfect mitigation can be learned. As such, the ansatz used in CGVD can be used to perfectly mitigate global depolarizing noise using a near-Clifford training set.

It is straightforward to extend this analysis to involve the ansatz including multiple noise levels. Notice that under the action of global depolarizing noise defined in~\eqref{eq:dep_noise_channel}, if this channel acts $j$ times throughout the dynamics we can write
\begin{equation}
    \rho_{j} \longrightarrow (1-\lambda^{j})\rho +\lambda^{j} \frac{\mathbb{1}}{2^Q}.
    \label{eq:dep_noise_channel_Multi}
\end{equation}

Now let us consider UNITED ansatz that combines observables mitigated with VD using various numbers of copies at various noise levels. This ansatz can be written as follows
\begin{align}
    \mu''_{M_{\rm max}, n} = \sum_{M=1}^{M_{\rm max}} \sum_{j = 1}^{n} d_{M, j} \hat{\mu}^{\rm VD}_{M, j},
\label{eq:UNITED_ansatz_2}
\end{align}
where $\mu''_{M_{\rm max},n}$ is the mitigated observable, $d_{M, j}$ are the parameters fitted using least squares regression on the data generated by the near-Clifford VD training circuits with up to $M_{\rm max}$ copies, evaluated with $n+1$ different noise levels.

Working through the same analysis with this ansatz, we find that, for the effects of global depolarizing noise to be perfectly mitigated we require
\begin{equation}
    \sum_{M=1}^{M_{\rm max}} \sum_{j=1}^{n} d_{M, j} f_{M, j} = 1,  \quad \tr{[X]} = 0 \ ,
\end{equation}
where
\begin{equation}
    f_{M, j} = 1 - \frac{2^Q \lambda^{Mj}}{(2^Q- 1) \lambda^{Mj} + ( 2^Q - \lambda^{j}(2^Q - 1 ))^M}.
\end{equation}
Therefore, we can conclude that both CGVD and UNITED can perfectly mitigate the effect of global depolarizing noise on some observable of interest.

The assumption $\tr{[X]}=0$ is not restrictive due to the common decomposition of operators into Pauli terms (which have zero trace) and the identity. However, we can consider an ansatz which includes an additional constant term to be fitted. In which case~\eqref{eq:CGVD_ansatz_2} becomes
\begin{align}
    \mu'_{M_{\rm max}} = \sum_{M=1}^{M_{\rm max}} b_{M} \hat{\mu}^{\rm VD}_{M} + b_0,
\label{eq:CGVD_ansatz_3}
\end{align}
Thus, \eqref{eq:perf_Mit_CGVD} becomes
\begin{equation}
    \mu'_{M_{\rm max}} = \sum_{M=1}^{M_{\rm max}} b_{M} \big( f_{M} \tr{[\rho X]} + \frac{1 - f_M}{2^Q}\tr{[X]}\big) +b_0.
\label{eq:perf_Mit_CGVD_2}
\end{equation}
Therefore, including a constant term in the ansatz, to perfectly mitigate the effect of depolarizing noise we require
\begin{equation}
    \sum_{M=1}^{M_{\rm max}} b_{M} f_{M} = 1,  \quad \sum_{M=1}^{M_{\rm max}} \frac{1-b_{M}}{2^Q} \tr{[X]} = b_0 \ .
\end{equation}
Including a constant term in the UNITED ansatz leads to a similar conclusion. Therefore, one can see that global depolarizing noise can be perfectly mitigated using near-Clifford circuits to train the parameters of the CGVD and UNITED ansatzes.

\section{Effects of the controlled derangement noise.}
\label{app:swap_noise}

\begin{figure}[t]
\includegraphics[width=1.\columnwidth]{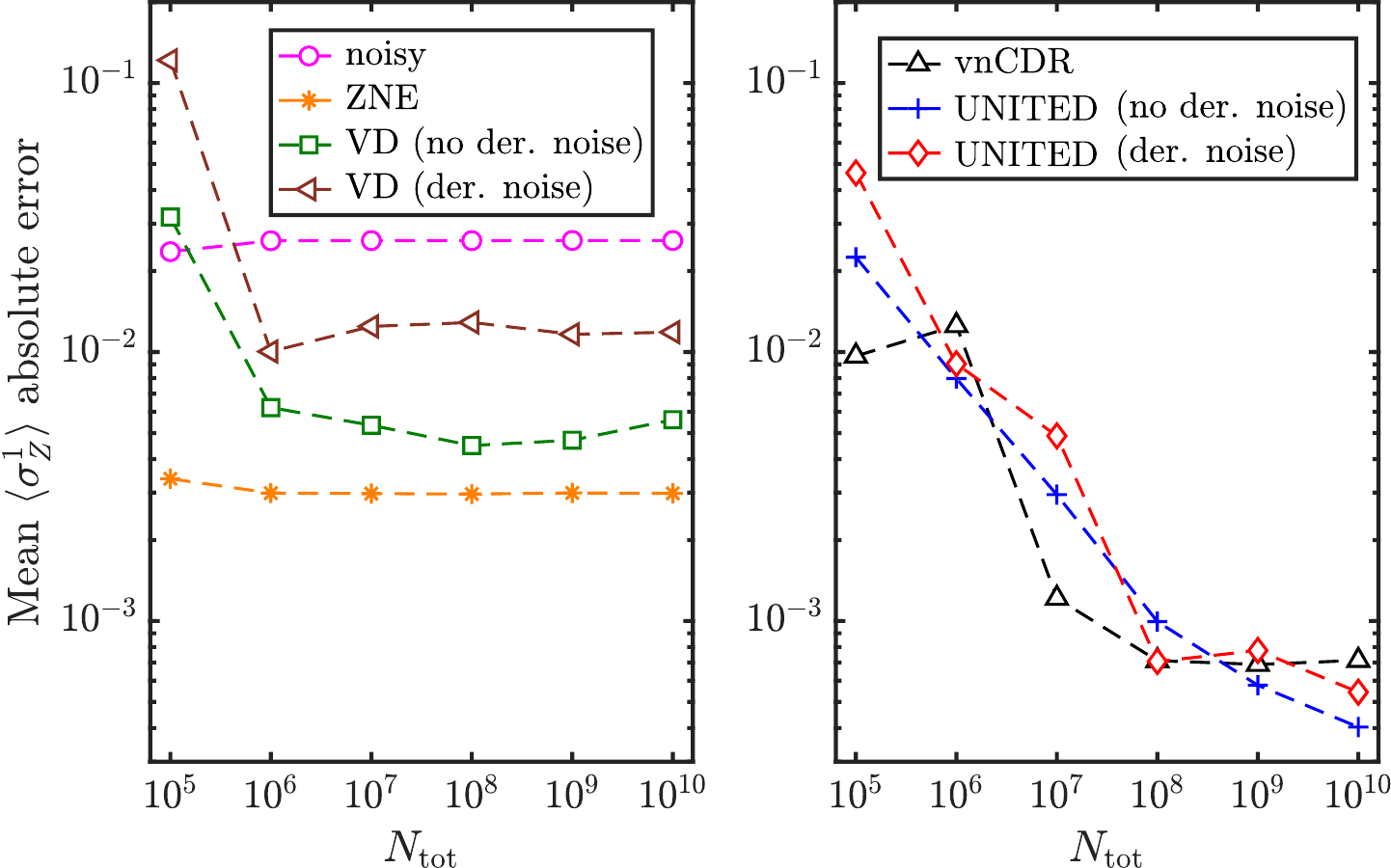}
\caption{\textbf{Comparing VD and UNITED error mitigation with (the brown and the red curves)  and  without controlled derangement noise (the green and the blue curves).} We perform the comparison for the random quantum circuits with $Q=4$ and $L=5$. We plot an error of mitigated and noisy  $\langle \sigma_Z^1 \rangle$ averaged over $30$ instances of the  random quantum circuits. The derangement noise affects VD and UNITED implementations decreasing the performance of VD and UNITED for large $N_{\rm tot}$. Nevertheless, even with the derangement noise, we obtain  a factor of $2.5$ improvement for VD with $N_{\rm tot} \ge 10^6$ and factor $1.2$ improvement of UNITED over vnCDR for the largest considered $N_{\rm tot} = 10^{10}$. For reference, we show also the noisy, ZNE, and vnCDR results. Here, ZNE is performed using linear extrapolation with noise levels $c=\{1,2\}$.  VD uses $M=3$ copies. We obtain the vnCDR correction using  noise levels $c=\{1,2,3\}$  and training circuits with $N=30$ $R_Z$ non-Clifford gates left intact. In the case of UNITED we have  $c=\{1,2,3\}$, $M_{\rm max}=3$, and $N=30$.}
\label{fig:swap_noise}
\end{figure}

\begin{figure*}[t!]
\centering
\includegraphics[width=0.6\textwidth]{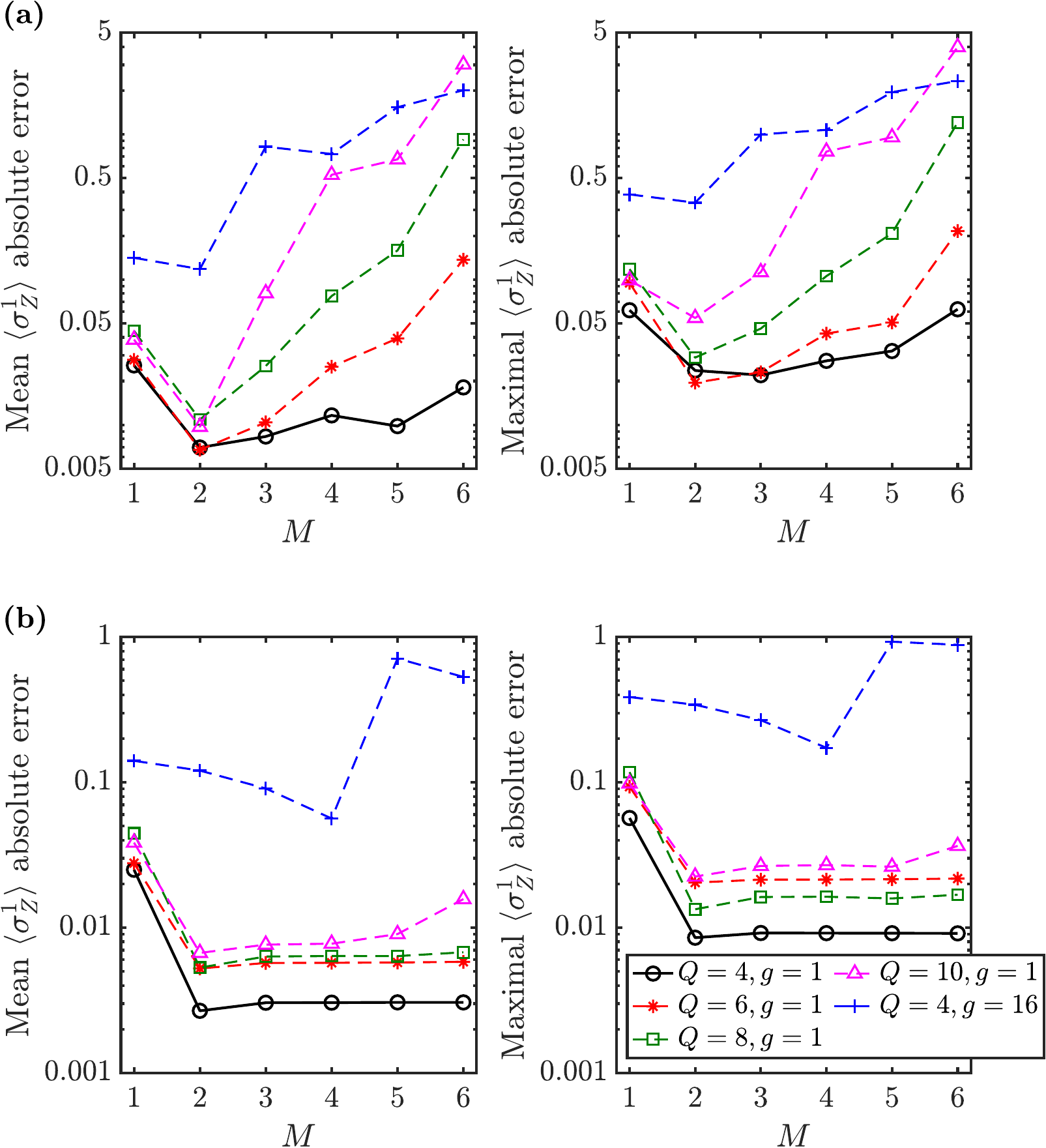}
\caption{\textbf{Convergence of VD estimates with increasing $M$.}  In (a) results for $N_{\rm tot}=10^5$  and in (b) results for $N_{\rm tot}=10^{10}$ plotted versus number of VD copies $M$. For reference purposes, we include single-copy noisy results as  $M=1$.  The plots show that   for $g=1$ $M=2$ is optimal while for $Q=4$, $g=16$ the optimal $M$ increases with increasing $N_{\rm tot}$.     
The left plots show the mean values while the right plots show the maximal values.} 
\label{fig:conv_M}
\end{figure*}

VD implementation on real hardware requires performing a controlled derangement which can be decomposed into controlled swap operations~\cite{koczor2020exponential}. In the main text we neglect the effects of noise in controlled swap implementations as the cost of classical simulation of the noisy controlled derangement scales as $2^{4Q+2}$ instead of the $2^{2Q}$ for the noiseless derangement simulation~\cite{czarnik2021qubit}. Here we examine the effects of the noisy derangement for $Q=4$ random quantum circuits. We perform the simulation by decomposing a controlled swap gate into two CNOTs and a Toffoli gate as described in~\cite{smolin1996five}. Subsequently, we decompose Toffoli gates and CNOTs to the native gates using  decomposition from~\cite{maslov2017basic}.    

In Figure~\ref{fig:swap_noise} we compare results for $Q=4$, $L=5$ random quantum circuits simulated with and without the derangement noise. We analyze error  $\langle \sigma^1_Z \rangle$  average over $30$ instances.  We find that the derangement noise reduces the performance ceiling of VD mitigation and the performance of UNITED for large $N_{\rm tot}=10^9,10^{10}$. Nevertheless, we still find a factor of $2.5$ VD improvement over the noisy simulations and a factor of $1.2$ UNITED improvement over vnCDR for $N_{\rm tot}=10^{10}$. 
These results indicate that UNITED can be useful in the presence of the derangement noise even without derangement noise error mitigation. 

Further improvements of UNITED and VD performance can be obtained by optimizing the decomposition of the controlled derangement circuit to  native gates~\cite{koczor2020exponential} and by using further error mitigation of controlled derangement noise~\cite{koczor2020exponential}. Another interesting direction is VD implementations based on state verification instead of controlled derangement~\cite{huo2021dual,cai2021resourceefficient}. 

\section{Optimal parameters for Virtual Distillation}
\label{app:VD_copy_analysis}

\begin{figure*}[t!]
\centering
\includegraphics[width=0.799\textwidth]{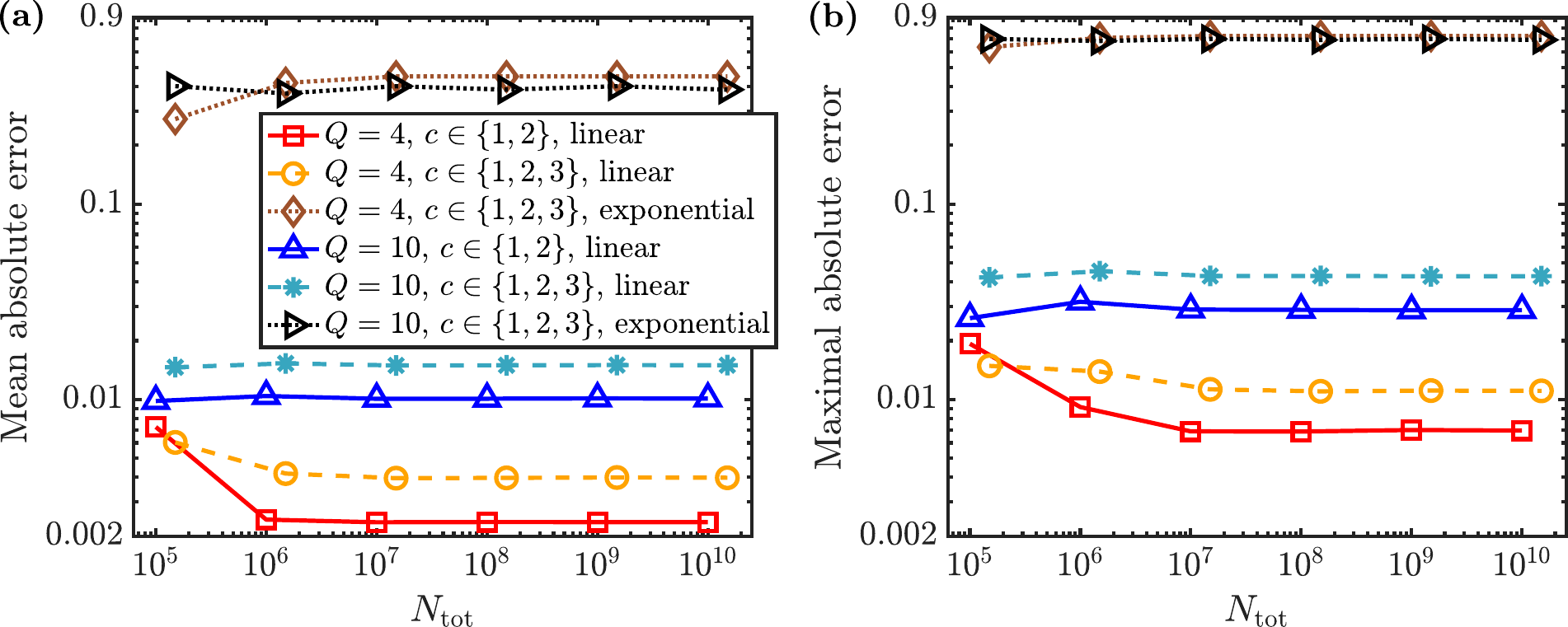}
\caption{\textbf{Convergence of  different ZNE variants  applied to RQCs.} Mean and maximal absolute  errors of the expectation value of  $ \sigma_Z^1 $ plotted versus the shot budget.  We use $30$ instances of random quantum circuits with $Q=4,10$ and $L=Q$. The red and blue curves are obtained with two noise levels $c\in\{1,2\}$ and a linear extrapolation while the orange and magenta curves are obtained with the  linear extrapolation and three noise levels $c\in\{1,2,3\}$.  The brown and the black curves are obtained with an exponential extrapolation and $c\in\{1,2,3\}$. }
\label{fig:ZNE_levels}
\end{figure*}

Here we  systematically explore the convergence of VD estimates with an increasing number of copies $M$. In Figure~\ref{fig:conv_M} we plot the mean and maximal errors of VD $\langle \sigma_Z^1 \rangle$ estimate for $30$ random quantum circuits and $M=2-6$. In this figure, we show results for  the smallest $N_{\rm tot}=10^5$ and the largest   $N_{\rm tot}=10^{10}$ considered here.  We find that $Q=4-10$ and $g=1$, $M=2$ typically gives optimal results for both shot budgets. We note that for $Q=4-10$, $g=1$ and $N_{\rm tot}=10^{10}$, $M=2-5$ VD gives very similar results indicating that $M=2$ is large enough to reach the noise floor. Our findings agree with~\cite{koczor2021dominant}, which argues that, with the exception of highly noisy circuits, $M\le3$ is sufficient to reach the noise floor.    For noisier  $Q=4,g=16$ circuits we find that the optimal $M$ depends strongly  on $N_{\rm tot}$.  For $N_{\rm tot}=10^5$, $M=2$ is optimal but it gives on average a small factor of $1.2$ improvement  over the noisy results.  For $N_{\rm tot}=10^{10}$, $M=4$ is optimal giving a factor of $2.5$ improvement. Based on the obtained results, we use $M=2$ in the main text as it is optimal for most circuits. We note that although with $M=4$ VD performance can be  improved for $Q=4,g=16$  and large $N_{\rm tot}$, this improvement is not large enough to outperform vnCDR and UNITED.  

\section{Comparison of  ZNE implementations}
\label{app:ZNE_levels}

For ZNE we compare the mitigation performance when using two or three noise levels, $c\in\set{1,2,3}$ and $c\in\set{1,2}$ respectively. For three noise levels, we apply both linear and exponential extrapolations, while the data coming from two noise levels only allows us to perform a linear extrapolation. 
We benchmark the extrapolation approaches when mitigating observables produced by RQCs and taking $N_{\rm tot} = 10^5, 10^6, \dots, 10^{10}$. For $L=Q$ and $Q=4,6,8,10$ we find that with the exception of $Q=4, N_{\rm tot}=10^{5}$ using two noise levels leads to better mitigation quality. The exponential extrapolation results in an order of magnitude worse results than the linear one. 
The results for $Q=4,10$ are shown in Figure~\ref{fig:ZNE_levels}.
For $L=16Q$  and  $Q=4,6,8$, we find that both variants of the linear extrapolation ZNE lead to similar mitigation quality.  The exponential extrapolation again produces an order of magnitude worse results than the linear extrapolation.
 
\section{Clustering of the  near-Clifford  circuits }
\label{app:clustering}

It is a well-known property of near-Clifford circuits that exact expectation values of Pauli strings cluster around~$0$. This clustering negatively impacts the shot-efficiency of learning-based error mitigation methods using near-Clifford circuits as training circuits~\cite{czarnik2022improving}. Here we demonstrate that the clustering does indeed occur for the training sets used to mitigate the random quantum circuit expectation values in Section~\ref{sec:RQC result}.

\begin{figure}[t!]
\includegraphics[width=0.99\columnwidth]{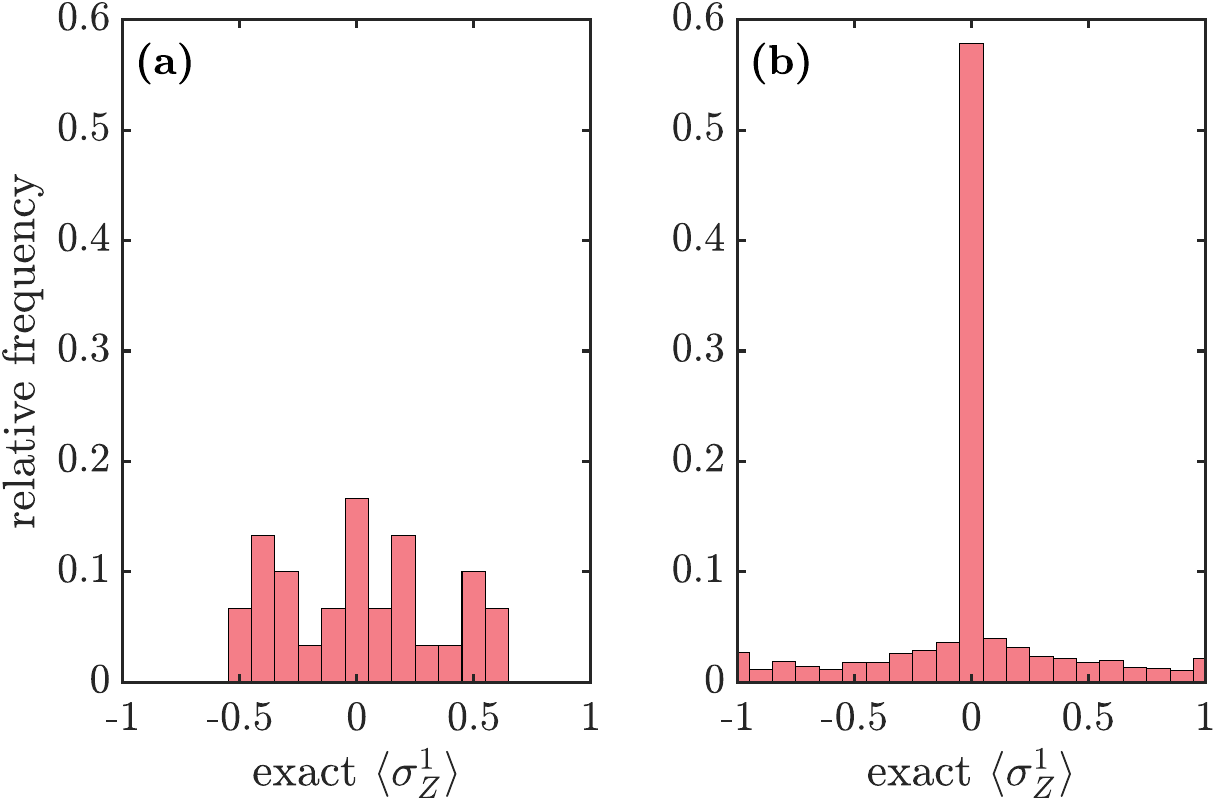}
\caption{\textbf{Clustering  of the near-Clifford circuits.}  In (a) we plot relative frequencies of different exact expectation values of the mitigated observable in a sample of $30$ random quantum circuits with $Q=4,L=4$. In (b) we show the same for  $3000$  near-Clifford   circuits  obtained by substitution of non-Clifford gates in the circuits analyzed in (a). For each random training circuit, $100$ near-Clifford circuits are generated. } 
\label{fig:hist}
\end{figure}

In Figure~\ref{fig:hist} we show relative frequencies of  exact expectation values of $\langle \sigma^1_Z \rangle$ 
for a  sample of  $30$  $Q=4,L=4$ random quantum circuits and  $3000$ near-Clifford circuits obtained by random substitutions of non-Clifford gates in these  random quantum circuits. For each circuit of interest, $100$ near-Clifford circuits were generated. We see strong clustering of the  expectation values for  the near-Clifford circuits despite the lack of this effect in the random quantum circuits used to generate them. The training circuits  used in Section~\ref{sec:RQC result} were then chosen by post-selecting on these near-Clifford circuits.

\begin{table*}[t!]
\centering
\small
\begin{tabular}{|c|c|c|c|c|c|c|}
\hline
$Q$ & $L$ & $|\langle \sigma^1_Z \rangle|_2$   &  $|\langle \sigma^1_Z \rangle|_3$   & $|\langle \sigma^1_Z \rangle|_2$   & $|\langle \sigma^1_Z \rangle|_3$ & 95th $|\langle \sigma^1_Z \rangle|$ percentile    \\
 &  &  RQC  &   RQC &  near-Clifford circuits   &  near-Clifford circuits & near-Clifford circuits \\
\hline
4 & 4 & 0.270 & 0.438 & 0  & 0.319 & 0.714\\
4 & 64 & 0.125 & 0.233 & 0.122 & 0.262 & 0.390 \\
6 & 6 & 0.185 & 0.300 & 0 &  0.081 & 0.584\\
6 & 96 & 0.108 & 0.146 & 0 & 0.074 & 0.212 \\
8 & 8 & 0.231 & 0.378 & 0 & 0 & 0.527 \\
8 & 128 & 0.030 & 0.063 & 0 & 0 &  0.067\\
10 & 10 & 0.184 & 0.265 & 0 & 0 & 0.363\\
\hline
\end{tabular}
\caption{\textbf{Clustering  of the near-Clifford circuits.} Here we gather medians $|\langle \sigma_Z^1\rangle |_2$  and third quartiles   $|\langle \sigma_Z^1\rangle |_3$ of $|\langle \sigma_Z^1\rangle |$, obtained for samples of $30$ random quantum circuits with various $Q$ and $L$ values  mitigated in Section~\ref{sec:RQC result}.  We also show $|\langle \sigma_Z^1\rangle |_2$,  $|\langle \sigma_Z^1\rangle |_3$  and 95\textsuperscript{th} percentile of $|\langle \sigma^1_Z \rangle|$ for $3000$ near-Clifford circuits obtained by substituting non-Clifford gates in the random quantum circuits. These circuits are then post-selected, choosing a subset that is used as the training circuits in UNITED and vnCDR in Section~\ref{sec:RQC result}. When the obtained quartiles are of the order of numerical precision ($10^{-15}$) or smaller we simply denoted it as $0$ in the table.       }
\label{tab:cluster}
\end{table*}

\begin{figure*}[t!]
\centering
\includegraphics[width=0.6\textwidth]{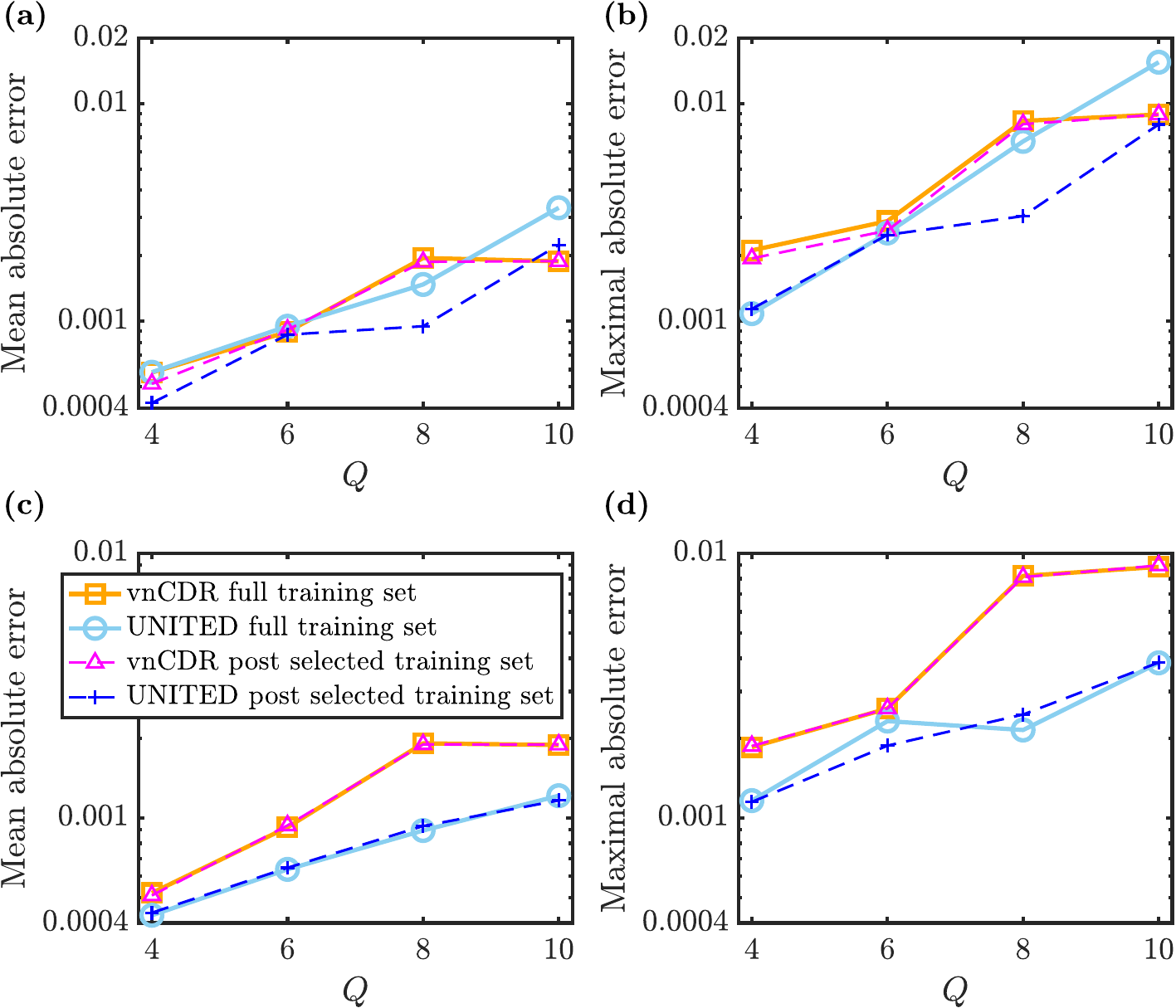}
\caption{{\bf Effects of training circuits post-selection on learning-based RQC error mitigation (UNITED and vnCDR).} Here we  compare the mean and maximal absolute errors of mitigated $\langle\sigma_Z^1\rangle$ for two methods of training circuits choice. First, we  the $N_t=100$  near-Clifford circuits obtained by random substitution of non-Clifford gates as the training circuits (the solid lines). Second, we post-select those circuits choosing the $N_t=50$ with the largest exact $\langle\sigma_Z^1\rangle$ as the training circuits (the dashed lines). The second strategy is used to obtain results in the main text.  Panels {\bf (a)} and {\bf (b)} show the results for the total error mitigation shot cost $N_{\rm tot}=10^{10}$. The results in the infinite $N_{\rm tot}$ limit are shown in {\bf (c)} and {\bf (d)}. Here we mitigate $30$ instances of $L=Q$ RQCs analyzed in Section~\ref{sec:RQC result}.  } 
\label{fig:training}
\end{figure*}

In Table~\ref{tab:cluster} we further analyze the clustering of the near-Clifford circuits. The table shows the median values $|\langle \sigma_Z^1 \rangle|_2$, and third quartiles  $|\langle \sigma_Z^1 \rangle|_3$ of the absolute value of  the observable of interest, obtained for all the random quantum circuits mitigated in Section~\ref{sec:RQC result} and all near-Clifford circuits generated before post-selection.  We see that $| \langle \sigma_Z^1 \rangle |_2$ and $|\langle \sigma_Z^1 \rangle |_3$ are decreasing slowly with increasing $Q$ for the random quantum circuits. For example, $Q=4, L=4$ the RQCs have $|\langle \sigma_Z^1 \rangle |_2 =0.270$ and  $|\langle \sigma_Z^1 \rangle |_2 =0.438$, while for the largest $Q=10$ ($L=10$) we obtain  $|\langle \sigma_Z^1 \rangle |_2 =0.184$ and  $|\langle \sigma_Z^1 \rangle|_2 =0.265$. Unlike the RQCs, the near-Clifford circuits are all strongly clustered around zero. Across all $Q$ and $L$ values mitigated in Section~\ref{sec:RQC result}, only $Q=4,L=64$ non-Clifford circuits have non-zero 
$|\langle \sigma_Z^1 \rangle |_2$. Furthermore,  $|\langle \sigma_Z^1 \rangle |_3=0$ for all $Q=8$ and $Q=10$ near-Clifford circuits.  Finally, we  gather in Table~\ref{tab:cluster} 95\textsuperscript{th} percentile of  $|\langle \sigma_Z^1 \rangle |$ for the near-Clifford circuits  to give information about tails of the near-Clifford  circuits distribution. We find that  it decreases with decreasing $Q$. Unlike in the case of $|\langle \sigma_Z^1 \rangle |_2$ and $|\langle \sigma_Z^1 \rangle |_3$, its decay is much faster for deep $L = 16Q$ circuits. For the deepest considered $Q=8,L=128$, it equals $0.067$, which indicates that nearly all generated near-Clifford circuits cluster around $0$.

These results show that the clustering of observables produced by near-Clifford circuits becomes stronger with increasing $Q$ much faster than in the case of RQCs. Consequently, they justify post-selection on the exact expectation values of  the training circuits employed while constructing the training circuits, which is described in detail in Section~\ref{sec:training_set_construction}. 
This technique improves the shot-efficiency of the error mitigation scheme as explored in more detail in~\cite{czarnik2022improving}.    
In Figure~\ref{fig:training} we compare  the mean and maximal absolute errors of the  mitigated observable obtained with and without post-selection on the training circuits. We show that indeed, the usage of the post-selected training set improves the quality of the error mitigation for a finite number of shots,  $N_{\rm tot}=10^{10}$, in both UNITED and vnCDR. We also compare the performance of both training set construction methods in the limit of an infinite number of shots. In that limit, both training sets with and without post-selection lead to a similar quality of error mitigation. The improvement of the  mitigation quality for $N_{\rm tot}=10^{10}$ is much  more pronounced for UNITED which  converges slower with  $N_{\rm tot}$ than vnCDR. This result  further demonstrates that  the post-selection is of particular importance when the quality of the  mitigation is limited by the shot budget  $N_{\rm tot}$.

\section{Details of  the noise model}
\label{app:noise_Model}

Here we use a  noise model  based on  Trout \textit{et al. }\cite{trout2018simulating} and developed in~\cite{cincio2021machine}. The noisy quantum  gates $R_X$, $R_Y$, $R_Z$ and $XX$ are defined with noise channels  $\mathcal S_\Xi$, $\mathcal S_{XX}$, respectively, where $\Xi$ is one of $X$, $Y$ or $Z$:
\begin{align}
    \mathcal S_\Xi = &\ \mathcal W^{p_d}\circ\mathcal D^{p_{\textrm{dep}}}\circ\mathcal R_{\Xi}^{p_\alpha} \,,\nonumber \\
    \mathcal S_{XX} = & \left[\mathcal W_1^{p_{d_1}}\otimes \mathcal W_2^{p_{d_2}} \right]
      \circ \left[\mathcal D_1^{p_{\textrm{dep}}}\otimes \mathcal D_2^{p_{\textrm{dep}}} \right]
      \circ\mathcal H^{p_\textrm{xx}}\circ\mathcal H^{p_h}.
\end{align}

Here, $\mathcal R_{\Xi}^{p_\alpha}(\rho) = (1-p_\alpha)\rho+p_\alpha \Xi \rho \Xi$, with $\Xi$ being one of $\sigma_{X}$, $\sigma_{Y}$ or $\sigma_{Z}$, which represent imprecision in the angle of rotation about the $X,Y,Z$ axes, respectively. $\mathcal D^{p_{\textrm{dep}}} $ is the local  depolarizing channel with the error rate $p^{\textrm{dep}}$, defined as $\mathcal D^p: \rho\mapsto   (1-p)\rho+ \frac{p}{3} ( \sigma_X\rho \sigma_X+ \sigma_Y\rho \sigma_Y+ \sigma_Z\rho \sigma_Z)$. $\mathcal W^{p_d}:\rho\mapsto(1-p_d)\rho+\sigma_Z\rho \sigma_Z$ is the dephasing channel, and $\mathcal H^p:\rho \mapsto(1-p)\rho +p  (\sigma_X \otimes \sigma_X) \rho  (\sigma_X \otimes \sigma_X)$ is a two-qubit channel that represents both an imprecise rotation when $p=p_{\rm xx}$ and ion heating when $p=p_h$.  Furthermore, we apply $\mathcal D^{p_{\textrm{idle}}}$ to simulate idling noise. The error rates were chosen as
\begin{align}
p_d&= 1.5\cdot10^{-4}, \quad   p_{\textrm{dep}}=  8\cdot10^{-4}, \nonumber  \\
p_{d_1}&=p_{d_2}=  7.5\cdot10^{-4}, \quad   p_\alpha= 1\cdot10^{-4} ,  \\
p_{\rm xx}&= 1\cdot10^{-3}, \quad            p_h=  1.25 \cdot10^{-3},\nonumber \\
p_{\textrm{idle}}&=8\cdot10^{-4}.\nonumber
\end{align}
We neglect measurement error in our simulations as it can be mitigated by specialized techniques~\cite{hamilton2020scalable, bravyi2021mitigating}.

\section{Compilation of the QAOA circuits to the native trapped ion gate set}

\label{app:comp}

In Section \ref{sec:qaoa} we mitigate expectation values of terms of the Max-Cut Hamiltonian (\ref{eq:maxcut}) for the QAOA ansatz (\ref{eq:qaoa})  $\langle\Psi(\vec{\alpha},\vec{\beta}\rangle)|\sigma_Z^i \sigma_Z^j|\Psi(\vec{\alpha},\vec{\beta})\rangle$.  To evaluate  them with the native trapped ion gate set  we notice  that 
\begin{equation}
 \langle \Psi(\vec{\alpha},\vec{\beta})  |  \sigma_Z^i \sigma_Z^j |   \Psi(\vec{\alpha},\vec{\beta}) \rangle  =  \langle \Psi'(\vec{\alpha},\vec{\beta})  |    \sigma_X^i \sigma_X^j |   \Psi'(\vec{\alpha},\vec{\beta}) \rangle,
\end{equation}
where, 
\begin{align}
|\Psi'(\vec{\alpha},\vec{\beta})\rangle &= e^{-i \beta_L H_M'} e^{-i \alpha_L H_P'} \dots \nonumber \\
&\dots e^{-i \beta_2 H_M'} e^{-i \alpha_2 H_P'} e^{-i \beta_1 H_M'} e^{-i \alpha_1 H_P'} |\vec{0} \rangle,
\end{align}
and $H_P' = \sum_{(i,j) \in E} \sigma_X^i \sigma_X^j$,  
$H_M' = \sum_{i=1}^Q \sigma_Z^i$.  
We have $e^{-i \alpha H_P'}=\prod_{(i,j)\in E} e^{-i \alpha \sigma_X^i \sigma_X^j} = \prod_{(i,j)\in E} XX(\alpha)$ and $e^{-i \beta H_M'} = \prod_{i=1,\dots,Q} e^{-i \beta \sigma_Z^i} = \prod_{i=1,\dots,Q} R_Z(\beta)$. That gives us the decomposition of $|\Psi'(\vec{\alpha},\vec{\beta})\rangle$ to the native gate set of a trapped-ion quantum computer.  

We note that $H_P'$ and $H_M'$ can be obtained from  the original problem Hamiltonian $H_P$ and   mixer Hamiltonian  $H_M$, respectively,  by a rotation  of the computational basis $U = \bigotimes_{i=1,\dots,Q} e^{-i (\pi/4) \sigma_Y^i}$, 
$H_P' = U  H_P U^{\dag}$, 
$H_M' = UH_M U^{\dag}$.
Therefore, $|\Psi(\vec{\alpha},\vec{\beta})\rangle$  and  $|\Psi'(\vec{\alpha},\vec{\beta})\rangle$ are also related by $U$, 
$|\Psi'(\vec{\alpha},\vec{\beta})\rangle = U |\Psi(\vec{\alpha},\vec{\beta})\rangle$.

\end{document}